\documentclass[11pt,sort&compress]{elsarticle}
\usepackage[paper=a4paper,top=24mm, bottom=26mm, left=26mm, right=26mm]{geometry}
\makeatletter
\def\ps@pprintTitle{%
 \let\@oddhead\@empty
 \let\@evenhead\@empty
 \def\@oddfoot{\centerline{\thepage}}%
 \let\@evenfoot\@oddfoot}
\makeatother
\usepackage{natbib}
\usepackage[hyperref]{xcolor}
\definecolor{darkgreen}{rgb}{0.01, 0.75, 0.24}
\definecolor{darkblue}{HTML}{2B66D3}
\usepackage[colorlinks=true,
            linkcolor=darkblue,
            urlcolor=darkblue,
            citecolor=darkblue,linkcolor=darkblue,hyperfootnotes=true]{hyperref}
            
\renewcommand\thefootnote{\textcolor{darkblue}{\arabic{footnote}}}            
\usepackage{amsmath}
\usepackage{amsfonts}
\usepackage{slashed}
\usepackage{accents}
\usepackage{amssymb}
\usepackage{mathrsfs}
\usepackage{mathtools}
\usepackage[utf8]{inputenc}
\usepackage[T1]{fontenc}
\let\oldbibliography\thebibliography
\renewcommand{\thebibliography}[1]{%
  \oldbibliography{#1}%
  \setlength{\itemsep}{1.4pt}%
}

\usepackage[cal=boondox,calscaled=1]{mathalfa}
\DeclareMathAlphabet{\bbvar}{U}{BOONDOX-ds}{m}{n}

\usepackage{psfrag}
\usepackage{pstool}
\usepackage{caption}

\usepackage{tabularx}
\usepackage{multirow}
\usepackage{pbox}
\usepackage{graphicx}
\newcommand{\q}[1]{`#1'\,}  
\newcommand{\utilde}[1]{\underaccent{\tilde}{#1}}
\newcommand{\di}{\mathrm{d}}
\usepackage{tensor}
\newcommand{\ou}[3]{\tensor{#1}{^{#2}_{#3}}}
\newcommand{\uo}[3]{\tensor{#1}{_{#2}^{#3}}}

\newcommand{\I}{\mathrm{i}} 
\newcommand{\E}{\mathrm{e}} 
\newcommand{\CC}{\mathrm{cc.}} 
\newcommand{\C}{\mathbb{C}}

\newcommand{\R}{\mathbb{R}}
\newcommand{\Z}{\mathbb{Z}}

\newcommand{\eref}[1]{(\ref{#1})}

\DeclareMathAlphabet{\bbgreek}{U}{bbold}{m}{n}

\newcommand{\mtext}[1]{\text{\it #1}}
\usepackage{tikz-cd}

\newcommand{\sh}{\operatorname{sh}}
\newcommand{\ch}{\operatorname{ch}}

\newcommand\vpm{\mathbin{\vcenter{\hbox{
  \oalign{\hfil$\scriptstyle+$\hfil\cr
          \noalign{\kern-.3ex}
          $\scriptscriptstyle({-})$\cr}}}}}
\DeclareMathAlphabet{\sfit}{OT1}{fos}{sb}{it}
\DeclareMathAlphabet{\mathsf}{OT1}{fos}{sb}{n}

\definecolor{darkgreen}{rgb}{0.01, 0.75, 0.24}
\definecolor{darkblue}{HTML}{2B66D3}

\usepackage[multiple, flushmargin]{footmisc}
\let\originalleft\left
\let\originalright\right
\renewcommand{\left}{\mathopen{}\mathclose\bgroup\originalleft}
\renewcommand{\right}{\aftergroup\egroup\originalright}

\newcommand{\dbarvar}{{\mathrm{d}\mkern-7.5mu\lower.18ex\hbox{$\textasciitilde$}\mkern-1.5mu}}

\renewcommand{\emph}[1]{{\it #1}}

\begin{document}

\begin{abstract}
\noindent We present a non-perturbative quantization of gravitational null initial data. Our starting point is the characteristic null initial problem for tetradic gravity with a parity-odd Holst term in the bulk. After a basic review about the resulting Carrollian boundary field theory, we introduce a specific class of impulsive radiative data. This class is defined for a specific choice of relational clock. The clock is chosen in such a way that the shear of the null boundary follows the profile of a step function. The angular dependence is arbitrary. Next, we solve the residual constraints, which are the Raychaudhuri equation and a Carrollian transport equation for an $SL(2,\R)$ holonomy. We show that the resulting submanifold in phase space is symplectic. Along each null generator, we end up with a simple mechanical system. The quantization of this system is straightforward. Our basic strategy is to start from an auxiliary Hilbert space with constraints.  The physical Hilbert space is the kernel of a constraint, which is a combination of ladder operators. The constraint and its hermitian conjugate are second-class. 
Solving the constraint amounts to imposing a simple recursion relation for physical states. On the resulting physical Hilbert space, the $SL(2,\R)$ Casimir is a Dirac observable. This observable determines the spectrum of the two radiative modes. The area at the initial and final cross sections are Dirac observables as well. They have a discrete spectrum, which agrees with earlier results on this topic.  
\end{abstract}
\title{Quantum geometry of the null cone}
\author{Wolfgang Wieland\footnote{wolfgang.wieland@fau.de}}
\address{Institute for Quantum Gravity, Theoretical Physics III, Department of Physics\\Friedrich-Alexander-Universität Erlangen-Nürnberg, Staudtstra\ss e 7, 91052 Erlangen, Germany\\{\vspace{0.5em}\normalfont 30 January 2024}
}

\maketitle
{\vspace{-1.2em}
\hypersetup{
  linkcolor=black
}
{\tableofcontents}

\hypersetup{
  linkcolor=darkblue,
  urlcolor=darkblue,
  citecolor=darkblue
}
\begin{center}{\noindent\rule{\linewidth}{0.4pt}}\end{center}\newpage\renewcommand\thefootnote{\textcolor{darkblue}{\arabic{footnote}}}   

\section{Introduction}\noindent

\noindent If gravity is quantum, we will find ourselves in a quantum spacetime geometry. What does it mean to live in such a world? What are the operators and eigenvalues that characterize the quantum nature of space and time? One of the main difficulties to answer these questions is the issue of general covariance. In a background invariant theory, geometry can  be described only relationally \cite{Rovelli_1991,partobs,Dittrich:2004cb}. This implies the existence of vast gauge redundancies. Complementary variables, such as the three-metric 
and Arnowitt--Deser--Misner (ADM) momentum  can 
provide only a redundant description of the physical state \cite{ADMEnergy,TEITELBOIM1973542}. 
If we start from some initial data on a partial Cauchy surface $\mathcal{B}$, we can always deform the surface into a new hypersurface $\mathcal{B}'$, while keeping fixed the corners $\mathcal{C}=\partial\mathcal{B}$, to obtain a new but gauge equivalent representation of the same physical state. For each gauge redundancy there is a generator, which vanishes as a constraint. To speak about observables, we need to impose the constraints and find the quantum numbers that characterize the physical, i.e.\ gauge invariant, data. 
There are different strategies to make this possible. One viewpoint is to attack the problem face-on. This is the canonical approach, in which one solves the constraints on a suitable kinematical Hilbert space \cite{thiemann, qsd,LOSTtheorem,Ashtekar:1994mh}. 
In here, following up earlier research on the subject \cite{Lehner:2016vdi,Parattu:2015gga,Donnelly:2016auv,Hopfmuller:2016scf,Freidel:2022vjq,Freidel:2020xyx,Freidel:2020svx,Chandrasekaran:2021hxc, Chandrasekaran:2019ewn,Ciambelli:2023mir,Reisenberger:2018xkn,Fuchs:2017jyk,Reisenberger:2012zq,AndradeeSilva:2022iic,Wieland:2017cmf,Wieland:2017zkf,Wieland:2019hkz,Wieland:2020gno,Wieland:2021vef,wieland:nulldefects}, we will argue for a slightly different viewpoint. Roughly speaking, the idea is as follows. Suppose, we had a quantum theory of gravity where there is a representation of the hypersurface deformation algebra. Consider a kinematical state $\rho_{\mathcal{B}}$ on a partial Cauchy surface $\mathcal{B}$, e.g.\ a three-ball with boundary $S^2$. Physical states are then equivalence classes $[\rho_{\mathcal{B}}]$ under hypersurface deformations. If we now take a diffeomorphism and deform $\mathcal{B}$ into a different three-surface $\mathcal{B}'$, such that the corner $\partial\mathcal{B}=\partial\mathcal{B}'$ is held fixed and $\mathcal{B}'$ lies in the future Cauchy development of $\mathcal{B}$, we obtain a new and gauge equivalent state $\rho_{\mathcal{B}'}\sim \rho_{\mathcal{B}}$. If we take this deformation to its natural extreme, we  obtain a state that lies at the boundary of the future Cauchy development of $\mathcal{B}$, which is a lightlike three-surface $\mathcal{N}$. If we now reverse the logic, we can start out from a quantum state at the null boundary $\mathcal{N}$ and take it as the unique representative (modulo residual diffeomorphisms) that defines the entire equivalence class $[\rho_{\mathcal{B}}]=:\rho_{\mathcal{N}}$. There are still residual gauge symmetries, but they are rather mild. This is so, because there is more structure available. A null surface is ruled by its light rays, points on different null rays are spacelike separated and the residual gauge redundancies break down to angle dependent reparametrizations of the null generators.\smallskip

Still, we may expect a preservation of difficulty. Further simplifying assumptions may be useful. In the following, we restrict ourselves to a certain class of \emph{impulsive null initial data} \cite{Aichelburg1971,Balasin:2007gh,Luk:2012hi,Griffiths:1991zp}, in which the shear is piecewise constant along the null generators. Each null generator, can then only carry countably many degrees of freedom. There are infinitely many degrees of freedom along the angular directions, but the constraints or transport equations do not couple neighbouring light rays, which are spacelike separated. Thus, we are dealing with a perhaps large but finite-dimensional Hamiltonian system on each light ray.  We are left with a standard quantum mechanical problem. 
 The practical usefulness of our approach lies also in certain infrared ambiguities that appear in perturbative gravity. If we take our description to  null infinity, each pulse has a soft graviton component, i.e.\ a spin-two excitation at $\mathcal{I}^\pm$, whose frequency is peaked around $\omega=0$. The resulting Fock states are 
no longer normalizable with respect to the naive $L^2$ inner product at $\mathcal{I}^\pm$, because of gravitational memory \cite{PhysRevLett.67.1486,Ashtekar:2018lor,Strominger:2017zoo,Strominger:2014pwa}. 
In our truncation, we replace a continuous profile by a large but finite sequence of pulses, and we will show below how each such pulse can be quantized at the full non-perturbative level. Thus, there is an implicit infrared cutoff built into our model.\smallskip

The paper splits into three main parts. In \hyperref[sec2]{Section 2}, we consider the {null boundary problem} for gravity in terms of spin connection variables with a parity violating Holst term \cite{holst} in the bulk. In addition, we will also present some new developments on the Carrollian geometry \cite{Freidel:2022vjq,Donnay:2019jiz,Ciambelli:2019lap,Ciambelli:2018ojf} of the spin bundle and clarify a few previous developments on charge quantization that we first made in \cite{Wieland:2017cmf,Wieland:2018ymr}. The third section, \hyperref[sec3]{Section 3}, provides a quick review about the quantisation of $T^\ast SL(2,\R)$ in terms of $SL(2,\R)$ boundary twistors. This quantization will prove useful at a later stage of this project when we will discuss how to quantise an entire sequence of pulses by gluing the states for the individual pulses across the interjacent corners. The main result is developed in \hyperref[sec4]{Section 4}, which consists of two parts. The first part is about the classical phase space. We take the covariant pre-symplectic potential and compute the pull-back to a submanifold in phase-space, where the shear of the null generators follows the profile of a step function. We impose the constraints and identify the Dirac observables at the full non-perturbative level. The second part takes the resulting finite-dimensional state space to the quantum level. We find a representation of the canonical Heisenberg relations and define the physical states of the system. We introduce the physical observables and discuss their spectra for a specific class of discrete representations of $SL(2,\R)$. Both the shear that characterizes the strength of the radiative data as well as the area of the cross sections of the null surface turn out to have a discrete spectrum. It is unclear at this point, whether the continuous series representations of $SL(2,\R)$ have any physical significance in our model. It is our expectation that they have not, for reasons explained in the paper. \smallskip

Our conventions follow previous articles on this topic, $a,b,c,\dots$ are abstract tensor indices, $\alpha,\beta,\gamma,\dots$ are internal Minkoski indices, $A,B,C,\dots$ and $\bar{A}, \bar{B}, \bar{C}$ are spinor indices ($\xi_A=\xi^B\epsilon_{BA}$, $\xi^A=\epsilon^{AB}\xi_B$ for $\epsilon_{AB}=-\epsilon_{BA}$, $\epsilon^{AC}\epsilon_{BC}=\delta^A_B$) transforming under the spin $(\tfrac{1}{2},0)$ and $(0,\tfrac{1}{2})$ representation of local $SL(2,\C)$ transformations. The configurations variable in the bulk are the cotetrad $e_\alpha\equiv e_{\alpha a}$ and the $\mathfrak{so}(1,3)$ spin connection $\ou{A}{\alpha}{\beta}\equiv\ou{A}{\alpha}{\beta a}$. Its curvature is $\nabla^2\equiv \ou{F}{\alpha}{\beta} =\di \ou{A}{\alpha}{\beta}+\ou{A}{\alpha}{\gamma}\wedge\ou{A}{\gamma}{\beta}$. The metric signature is $(-$$+$$+$$+)$.
\section{Boundary field theory and null symplectic structure}\label{sec2}
\subsection{Holst action in a causal domain}
\label{sec2.1}
\noindent In four spacetime dimensions, we can add a parity breaking term to the Hilbert--Palatini action, without changing the field equations in the interior \cite{holst,surholst,PhysRevD.105.064066,komplex1}. This freedom, which does not exist in higher dimensions, is responsible for a new coupling constant, namely the Barbero--Immirzi parameter $\gamma$ (w.l.g.\ $\gamma>0$), which can enter the action in the bulk. When there is no cosmological constant and all matter couplings are set to zero, the resulting \emph{Holst action} is
\begin{equation}
S_{\mathcal{M}}[A,e]=\frac{1}{16\pi G}\int_\mathcal{M}\Big[\operatorname{\ast}(e_\alpha\wedge e_\beta)\wedge F^{\alpha\beta}[A]-\frac{1}{\gamma}e_\alpha\wedge e_\beta\wedge F^{\alpha\beta}[A]\Big],\label{actnblk0}
\end{equation}
where $\ast(e_\alpha\wedge e_\beta)=\tfrac{1}{2}\uo{\epsilon}{\alpha\beta}{\alpha'\beta'}e_{\alpha'}\wedge e_{\beta'}$ is the Hodge dual of the Pleba\'{n}ski two-form $\Sigma_{\alpha\beta}=e_\alpha\wedge e_\beta$, which is a two-form that takes values in the Lorentz bundle. On shell, both terms in the action vanish. The first term vanishes thanks to the vacuum Einstein equations $\ast F_{\alpha\beta}\wedge e^\beta=0$ and the second term vanishes thanks to the Bianchi identity $F_{\alpha\beta}\wedge e^\beta=0$, which is a consequence of the torsionless condition $\nabla e^\alpha=\di e^\alpha+\ou{A}{\alpha}{\beta}\wedge e^\beta$.\smallskip

In the following, we consider how the Barbero--Immirzi parameter effects the {null boundary problem} of classical and quantum general relativity. This is to say that we study the state space and algebra of observables derived from \eref{actnblk0} in a four-dimensional spacetime region $\mathcal{M}$ with a boundary $\partial\mathcal{M}=\mathcal{B}_+\cup\mathcal{N}\cup\mathcal{B}_-$ that contains a three-dimensional null boundary $\mathcal{N}$. The bottom and top of the boundary $\partial\mathcal{M}$ are space-like three-balls $\mathcal{B}_{\pm}$ that are connected to $\mathcal{N}$ at co-dimension two boundaries (corners) $\mathcal{C}_{\pm}=\partial\mathcal{B}_{\pm}$. For simplicity and definiteness of the problem, we assume that $\mathcal{C}_\pm$ have the topology of a two-sphere. In addition, the null portion of the boundary is viewed as an abstract topological manifold $\mathcal{N}=(-1,1)\times S^2$, where $\mathcal{C}_+=\{+1\}\times [S^2]^{-1}$ and  $\mathcal{C}_-=\{-1\}\times S^2$ are topological two-spheres. Any such null surface $\mathcal{N}$ is equipped with a projector $\pi:\mathcal{N}\rightarrow\mathcal{C}, \mathcal{N}\ni(u,q)\mapsto q\in\mathcal{C}$, which maps every light ray in $\mathcal{N}$ into a point on the base manifold. Accordingly, we can say that a vector field $l^a\in T\mathcal{N}$ is null, if its push-forward onto the base manifold vanishes, i.e.\ $\pi_\ast l^a=0$. This condition serves as a partial gauge fixing in which the direction of the light rays that are intrinsic to $\mathcal{N}$ is kept fixed once and for all. Apart from the boundary condition $u|_{\mathcal{C}_\pm}=\pm 1$, the time parameter on the null boundary is completely arbitrary. In particular, it is not an affine or otherwise geometrically preferred null coordinate on $\mathcal{N}$.\smallskip

 At the co-dimension two corners, otherwise unphysical gauge modes become physical \cite{Freidel:2023bnj,Speranza:2017gxd,Balachandran:1994up,Barnich:2011mi, PhysRevD.51.632, Freidel:2021cjp, Freidel:2020xyx,Freidel:2020svx,Donnelly:2016auv,Wieland:2021vef,Freidel:2020ayo,Jacobson:2022gmo,Wieland:2020gno,Wieland:2023qzp}. This is a direct consequence of how the then necessary boundary conditions break the local gauge symmetries of general relativity in the bulk. Take a generic solution in the interior and consider the formal evolution from the initial values at $\mathcal{B_-}$ to $\mathcal{B_+}$ along some vector field $\xi^a$, which is null if we restrict it to $\mathcal{N}$. Since radiation can cross the null boundary, the resulting flow equations  can no longer be Hamiltonian. The system is dissipative. Additional radiative data must be fixed before we can integrate the field equations from $\mathcal{B_-}$ to $\mathcal{B_+}$. Since a generic four-dimensional diffeomorphism moves the null boundary relative to the fields in the bulk, the required boundary conditions break diffeomorphism invariance. Once a gauge symmetry is broken, we obtain a new additional physical mode. Since the symmetry is only broken at the boundary, the resulting physical phase space on $\mathcal{B}_\pm$ consist of radiative modes inside and additional boundary modes at $\mathcal{C}_\pm$. The boundary modes themselves split into edge modes $\lambda$, which serve as reference frames \cite{Carrozza:2021gju,Carrozza:2022xut,PhysRevD.108.106022} for \emph{large} gauge symmetries, and their conjugate momenta, which define the charges $Q_{\lambda}$ at the boundary.\footnote{Small gauge symmetries, on the other hand, have gauge parameters $\lambda$ that vanish at the corners.  They are mere gauge redundancies of the theory in the bulk.  Large gauge symmetries are not. }\smallskip
 
Although the Barbero--Immirzi parameter does not enter the field equations $\ast F_{\alpha\beta}\wedge e^\beta=0$ and $\nabla e^\alpha=0$ in the bulk, it can have observables consequences. It alters the symplectic structure, which determines the quantisation map. This alteration is subtle. It does not change the classical Poisson algebra of the two radiative modes at the null boundary \cite{Wieland:2021vef}. It only affects the boundary modes. In this way, the generators of boundary symmetries depend explicitly on $\gamma$. At the classical level, this amounts to a mere relabelling of canonical charges, which does not change the physics. 
Consider then the dictionary between the abstract gauge charges $\{Q_{\lambda}\}$ and more standard observables $\{O_{\alpha}\}$ such as mass, spin, angular momentum. If we start from the action \eref{actnblk0}, the map between $\{O_\alpha\}$ and  $\{Q_{\lambda}\}$ depends on $\gamma$. Hence there is some functional dependence $Q_{\lambda}(\gamma,\{{O}_\alpha\})$. As far as the classical observables are concerned, this dependence amounts to a  mere relabelling of the list of boundary charges. The numerical value, that each of the physical observables $\{{O}_\alpha\}$ has, can only depend on the metric $g_{ab}$ and its derivatives at the boundary, i.e\ ${O}_\alpha={O}_\alpha[g_{ab},\nabla_a,\mathcal{C}]$. Given a classical solution $g_{ab}$ of the field equations,  the observables $\{O_{\alpha}\}$ are independent of $\gamma$. At the quantum level, the situation is more subtle. Now, the logic is reversed. We start from the algebra of observables, schematically $[\hat{Q}_\lambda,\hat{Q}_{\lambda'}]=\hat{Q}_{[\lambda,\lambda']}$, and seek for a representation of the canonical commutation relations on a Hilbert space. Given such a representation, we can predict possible eigenvalues for the charges. It is only in a second step that we can then invert the map $Q_{\lambda}(\gamma,\{{O}_\alpha\})$ between the charges $\{Q_\lambda\}$ and the observables $\{O_\alpha\}$. The commutation relations for the boundary charges are determined  by 
their algebraic properties $\{\hat{Q}_\lambda\}$, which can be inferred from the symmetries they generate. Hence, there is no dependence on $\gamma$. All eigenvalues for the charges are independent of $\gamma$.  Yet, the map $Q_{\lambda}(\gamma,\{{O}_\alpha\})$ depends on $\gamma$. Thus, the Barbero--Immirzi parameter can enter the spectrum of physical observables $\{{O}_\alpha\}$.   Besides the boundary modes, there is also the radiative data. Since the charges must be consistent with certain flux-balance laws for the free radiative data, the radiative data will also be affected by the addition of $\gamma$. A similar mechanism appears in QED in the presence of a $\theta$-term in the action, which can alter the quantization condition between electric and magnetic charges \cite{Witten:1979ey}. For each different value of $\gamma$, we can expect, therefore, unitarily inequivalent representations of the canonical commutation relations. Depending on the value of $\gamma$, the quantisation of the radiative data will be different. Since $\gamma$ does not enter the classical Poisson algebra for the radiative data, this may sound very surprising at first.  But we will see below how this mechanism can play out in practice.

\subsection{Boundary conditions and boundary Carollian field theory}\label{sec2.2}
\noindent Depending on which boundary conditions we choose, we need to add different boundary terms to the action. In the following, we consider a class of conformal boundary conditions along the null boundary for the Holst action \eref{actnblk0} in the bulk. The treatment becomes particularly transparent in terms of self-dual (spinor-valued) two-forms. The action in the bulk \eref{actnblk0} can be now seen as a sum of selfdual and anti-selfdual parts,
\begin{equation}
S_{\mathcal{M}}[A,e]=\Big[\frac{\I}{8\pi\gamma G}(\gamma+\I)\int_\mathcal{M}\Sigma_{AB}\wedge F^{AB}\Big]+\CC,\label{actnblk1}
\end{equation}
where $\ou{F}{A}{B}=\di \ou{A}{A}{B}+\ou{A}{A}{C}\wedge\ou{A}{C}{B}$ is the curvature of the self-dual connection and $\Sigma_{AB}$ is the Pleba\'{n}ski two-form
\begin{equation}
\Sigma_{AB}=-\frac{1}{2}e_{A\bar{C}}\wedge\uo{e}{B}{\bar{C}},
\end{equation}
where $e_{AA'}$ are the tetrads in spinor notation.

A possible boundary term for conformal boundary conditions on a null surface for the Holst action was introduced in \cite{Wieland:2017zkf,Freidel:2023bnj}. The simplest way to understand the corresponding boundary Lagrangian is to start out from the pull-back to the null boundary of the Pleba\'{n}ski two-form in the bulk. A straightforward calculation gives
\begin{equation}
\varphi^\ast_{\mathcal{N}}\Sigma_{AB}=e_{(A}\wedge\bar{m}\,\ell_{B)},\label{glucond}
\end{equation}
where $\ell^A$ is a section of the spinor bundle over the null surface and $e_A$ is a spinor-valued one-form at the boundary. The complex-valued one-forms $m_a$, on the other hand, define the pull-back of the spacetime metric to the null boundary, 
\begin{equation}
\varphi^\ast g_{ab}=2m_{(a}\bar{m}_{b)}.
\end{equation}
The wedge product of $m$ and $\bar{m}$ gives the area two-form $\varepsilon$ of the null surface,
\begin{equation}
\varepsilon_{ab}=-2\I m_{[a}\bar{m}_{b]}.
\end{equation}
To extend the dyads $(m_a,\bar{m}_a)$ into a co-basis of $T^\ast\mathcal{N}$, we introduce a one-form
\begin{equation}
k=-N\di u,\label{kdef}
\end{equation}
where $u:\mathcal{N}\rightarrow(-1,1)$ is the time-coordinate, which  we introduced earlier, see \hyperref[sec2.1]{section 2.1}. We do not impose any specific condition on $u$ other that it increase monotonically along the null generators $\gamma_q=\pi^{-1}(q)$, $q\in\mathcal{C}$. The scalar function $N$, on the other hand, plays the role of a lapse function. Given such a triple $(k_a,m_a,\bar{m}_a)$, we then also have a corresponding dual basis, which is denoted by $(\ell^a,m^a,\bar{m}^a)$. It satisfies $k_a\ell^a=-1$ and $m_a\bar{m}^a=1$. All other contractions vanish. Therefore, $\ell^a$ is a null vector, i.e.\ $\pi_\ast\ell^a=0$.

The exterior derivatives of the one-forms $(m_a,\bar{m}_a)$ define spin connection coefficients intrinsic to $\mathcal{N}$,
\begin{equation}
\di m = -\Big(\tfrac{1}{2}\vartheta_{(\ell)}+\I\varphi_{(\ell)}\Big)k\wedge m -\sigma_{(\ell)}\,k\wedge\bar{m}+\I\,\gamma\,m\wedge\bar{m},
\end{equation}
where $\vartheta_{(\ell)}$ is the expansion of the null generators $\ell^a$ and $\sigma_{(\ell)}$ is the shear. The coefficients $\varphi_{(\ell)}$ and $\gamma$, on the other hand, can be rearranged into a $U(1)$ connection
\begin{equation}
\Gamma = k\varphi_{(\ell)}+\gamma\bar{m}+\bar{\gamma}m.\label{Gammadef}
\end{equation}
Consider then the components of the one-form $e_A$ with respect to the co-basis $(k_a,m_a,\bar{m}_a)$,
\begin{equation}
e_A\wedge \bar{m}=(\ell_A k-k_Am)\wedge m.
\end{equation}
On the null surface, the spinors $k^A$ and $\ell^A$ are linearly independent and satisfy $k_A\ell^A=1$. Hence they define a basis in the spinor bundle over $\mathcal{N}$. We can now take the pull back of the selfdual connection to $\mathcal{N}$, thereby introducing a covariant derivative\footnote{The corresponding connection, which contains depends on both extrinsic and intrinsic data at $\mathcal{N}$, is the null surface analogue of the self-dual Ashtekar connection \cite{newvariables} on a spatial hypersurface.} $D=\varphi^\ast_{\mathcal{N}}\nabla$, and obtain spin rotation coefficients. If the torsionless condition is satisfied, we find
\begin{align}
\ell_AD\ell^A&=-\frac{1}{2}\big(\vartheta_{(\ell)}m+\sigma_{(\ell)}\bar{m}\big),\label{expnssheardef}\\
k_AD\ell^A&=\frac{1}{2\I}\big(\Gamma+\I K\big),
\end{align}
where $\Gamma$ is the $U(1)$ connection \eref{Gammadef} and the one-form $K$ depends on only extrinsic data. It admits the decomposition
\begin{equation}
K=-\kappa_{(\ell)}k+\bar{\alpha} m+\alpha\bar{m},\eref{Kdef}
\end{equation}
where $\kappa_{(\ell)}$ is the affinity of the null generators $\ell^a$ and the coefficient $\alpha$ satisfies
\begin{equation}
\bar{\alpha}=-N^{-1}\bar{m}^a\partial_aN+\ell^bk_AD_bk^A.
\end{equation}
The components $m^bk_AD_bk^A=\tfrac{1}{2}\vartheta_{(k)}$ and $\bar{m}^bk_AD_bk^A=\bar{\sigma}_{(k)}$ are the shear and expansion of the transversal null direction $k^a\equiv \I k^A\bar{k}^{\bar{A}}$. In the following, these two transversal components play no further role in our analysis.\smallskip

Consider then the following shifted and complexified $U(1)$ connection $A$,
\begin{equation}
A = -\I\Gamma + K + \vartheta_{(\ell)}k.\label{Adef}
\end{equation}
Given this connection, it is then easy to show that the boundary spinors satisfy the following boundary equations along the null surface
\begin{align}
\bar{m}\wedge \big(D-\tfrac{1}{2}A\big)&=\tfrac{1}{2}\vartheta_{(\ell)}e^A\wedge\bar{m},\label{bndryEOM1}\\
\big(D+\tfrac{1}{2}A\wedge\big)(e_A\wedge\bar{m})&=0.\label{bndryEOM2}
\end{align}
Equations \eref{bndryEOM1} and \eref{bndryEOM2} determine Carollian transport equations for the boundary spinors $\ell^A$ and $e_A\operatorname{mod}\bar{m}$ along the null generators. A boundary action, whose equations of motion are \eref{bndryEOM1} and \eref{bndryEOM2} is then given by
\begin{equation}
S_{\mathcal{N}}[e_A,\ell^A,\ou{A}{A}{B}|m,A,\vartheta]=\frac{\I}{8\pi \gamma G}(\gamma+\I)\int_{\mathcal{N}}\Big[e_A\wedge\bar{m}\wedge(D-\tfrac{1}{2}A)\ell^A-\tfrac{\vartheta}{4}e_A\wedge e^A\wedge\bar{m}\Big]+\CC,\label{bndryactn}
\end{equation}
where $\vartheta=\bar{\vartheta}$ is real, while all other configuration variables are complex. The action assumes the standard form of Hamiltonian mechanics, in which the action is a difference of a symplectic term $\theta\propto e_A\wedge\bar{m}\wedge(D-\tfrac{1}{2}A)\ell^A$ and a  Hamiltonian  $h\propto \tfrac{\vartheta}{4}e_A\wedge e^A\wedge\bar{m}$, which is quadratic in the momenta $\pi_A=e_A\operatorname{mod}(\bar{m})$. In computing the saddle points of the action, we keep a gauge equivalence class of boundary fields $[m,A,\vartheta]$ fixed:
\begin{equation}
\textsl{g}=[m,A,\vartheta]/_\sim,\qquad \delta\textsl{g}=0.\label{bndrycond}
\end{equation}
The symmetries that define this equivalence class are
\begin{description}
\item[- \emph{Vertical diffeomorphisms:}] $[\varphi^\ast m,\varphi^\ast A,\vartheta\circ\varphi]\sim[m,A,\vartheta]$ for all $\varphi\in\mathrm{Diff}(\mathcal{N}:\mathcal{N}):\pi\circ\varphi=\mathrm{id}_{\mathcal{C}}$.
\item[- \emph{$U(1)$ transformations:}] $[\E^{\I\phi} m,A+\I\,\di\phi,\vartheta]\sim[m,A,\vartheta]$ for all $\phi:\mathcal{N}\rightarrow\R$.
\item[- \emph{Dilatations of the null normal:}] $[m,A+\di\lambda,\E^\lambda\vartheta]\sim[m,A,\vartheta]$ for all $\lambda:\mathcal{N}\rightarrow\R$.
\item[- \emph{Conformal rescalings:}] $[\E^\omega m,A,\E^\omega\vartheta]\sim[m,A,\vartheta]$ for all $\omega:\mathcal{N}\rightarrow\R$.
\item[- \emph{Shifts of the $U(1)_\C$ connection:}] $[m,A,\vartheta]\sim[m,A+\frac{\I}{\gamma+\I}\beta+\bar{\xi}_+m+\xi_-\bar{m},\vartheta]$ for all one-forms $\beta\in\Omega^1(\mathcal{N}:\R)$ and scalars $\xi_\pm:\mathcal{N}\rightarrow\C$. This condition removes all but just one of the $3\times 2=6$ real components of the abelian boundary connection $A\in\Omega^1(\mathcal{N}:\C)$. The residual component is essentially the affinity $\kappa_{(\ell)}$ shifted by terms proportional to $\vartheta_{(\ell)}$ and $\varphi_{(\ell)}$ that only depend on the intrinisic geometry of $\mathcal{N}$.
\end{description}
Notice that we have fixed earlier the null directions of $\mathcal{N}$ as an external background structure, i.e.\ we do not change how to embed the abstract light rays $\pi^{-1}(q), q\in\mathcal{C}$ into the null boundary $\mathcal{N}$. This implies that the one-form $m_a$ can only have two complex degrees of freedom, since $l^am_a=0$ for all $l^a\in T\mathcal{N}:\pi_\ast l^a=0$. The triple $[m,A,\vartheta]$ is thus specified by $4+6+1=11$ local parameters on $\mathcal{N}$. The equivalence relation removes nine of them. The remaining two degrees of freedom characterize the free initial data along $\mathcal{N}$.\smallskip

The coupled bulk plus boundary action is the sum $S=S_{\mathcal{M}}+S_{\mathcal{N}}$ of \eref{actnblk1} and \eref{bndryactn}. If we now vary the action and determine its stationary points for the conformal boundary conditions \eref{bndrycond}, we obtain the usual Einstein equations in the bulk and additional gluing conditions and boundary field equations. The boundary field equations \eref{bndryEOM1} and \eref{bndryEOM2} transport the boundary spinors $\ell^A$ and $e_A\wedge\bar{m}$ along the null generators. The gluing conditions \eref{glucond} link them to the fields in the bulk. If we then take variations of the action that violate the boundary conditions $\delta\textsl{g}\neq 0$, we obtain the symplectic structure. After some straightforward algebraic manipulations, we obtain the symplectic structure on the null surface $\mathcal{N}$, which is now given by
\begin{equation}
\Theta_{\mathcal{N}}=\frac{\I}{8\pi\gamma G}(\gamma+\I)\int_{\mathcal{N}}\Big[\bbvar{d}(k\wedge\bar{m})\wedge \ell_AD\ell^A+\I\varepsilon\wedge\bbvar{d}A\Big]+\CC,\label{Theta1}
\end{equation}
where $\bbvar{d}$ is the exterior derivative on the covariant phase space.\footnote{N.B. A perhaps much faster way to arrive at the same result is to start out from the symplectic potential $\Theta_{\mathcal{B}}=\frac{\I}{8\pi\gamma G}(\gamma+\I)\int_{\mathcal{B}}\Sigma_{AB}\wedge\bbvar{d}A^{AB}+\CC$ on a partial Cauchy surface $\mathcal{B}$, then take a formal limit in which $\mathcal{B}$ is sent to a null surface, and impose, by hand, that all the variations of the spin dyad vanish  $\bbvar{d}\ell^A=\bbvar{d}k^A=0$.}\smallskip

Let us close this section with a few comments on the residual diffeomorphism symmetries. The abstract null boundary $(\mathcal{N}=(-1,1)\times\mathcal{C},\pi)$ is equipped with a projector onto the corner $\mathcal{C}$, $\pi:\mathcal{N}\rightarrow\mathcal{C}$. For any $q\in\mathcal{C}$, the light rays $\gamma_q=\pi^{-1}(q)$ provide us with a background structure, which we keep fixed.\footnote{This can be justified relationally. To say that two events $x$ and $x'$ lie on the same null ray, $\pi(x)=\pi(x')$, is a coordinate invariant statement.} Thus, any variation of the null normal $\bbvar{d}\ell^a$ will be again proportional to $\ell^a$, $\bbvar{d}\ell^a\propto\ell^a$. This, in turn, implies that the diffeomorphism symmetries of the null surface are broken down to the semi-direct product of $\mathrm{Diff}(\mathcal{C})\ltimes \mathrm{Diff}(I)^{\mathcal{C}}$, where $\mathrm{Diff}(I)$ are the diffeomorphisms of the intervall $(-1,1)\ni u$. Of these, only the subgroup $\mathrm{Diff}(I)^{\mathcal{C}}$ generated by vertical vector fields $\xi^a\in T\mathcal{N}:\pi_\ast\xi^a=0$ that vanish at the two corners are redundant null directions of the pre-symplectic two-form $\Omega_{\mathcal{N}}=\bbvar{d}\Theta_{\mathcal{N}}$. Elements of the rotational subgroup $\mathrm{Diff}(\mathcal{C})$, on the other hand, do not generate gauge redundancies. They are true physical symmetries, i.e.\ motions on the physical phase space. The residual gauge redundancies are generated by a constraint, namely the Raychaudhuri equation
\begin{align}
0=\I&\varphi^\ast_{\mathcal{N}}\Big(F_{AB}\wedge\ou{e}{B}{\bar{A}}\ell^A\bar{\ell}^{\bar{A}}\Big)=(\ell_A D^2\ell^A)\wedge\bar{m}=\nonumber\\
&=\big(\di(\ell_AD\ell^A)-D\ell_A\wedge D\ell^A\big)\wedge\bar{m}=\big(\di(\ell_AD\ell^A)+\I(\Gamma+\I K)\wedge \ell_AD\ell^A\big)\wedge\bar{m}\nonumber=\\
&=\frac{1}{2}\Big(\ell^a\partial_a[\vartheta_{(\ell)}]-\kappa_{(\ell)}\vartheta_{(\ell)}+\frac{1}{2}\vartheta_{(\ell)}^2+2\sigma_{(\ell)}\bar{\sigma}_{(\ell)}\Big)k\wedge m\wedge\bar{m}.\label{Req0}
\end{align}

By keeping the null directions of $\mathcal{N}$ fixed, we reduced $\mathrm{Diff}(\mathcal{N})$ down  $\mathrm{Diff}(\mathcal{C})\ltimes\mathrm{Diff}(I)^{\mathcal{C}}$, of which only the normal subgroup of angle dependent reparametrizations of the null direction (forming the little group of $\mathcal{N}$, so to say) survive as redundant gauge symmetries (i.e.\ null directions of $\Omega_\mathcal{N}=\bbvar{d}\Theta_{\mathcal{N}}$). The elements of $\mathrm{Diff}(\mathcal{C})$, on the other hand, are genuine symmetries generated by boundary charges, see e.g.\ \cite{Barnich:2011mi,Odak:2023aa,Chandrasekaran:2018aop,Wieland:2021vef,Wieland:2017zkf}. At this point it is useful to compare our situation to the Hamiltonian analysis in terms of ADM variables, where there are four constraints on a spatial hypersurface $\Sigma$. There are the three vector constraints that generate $\mathrm{Diff}(\Sigma)$ and there is one additional scalar constraint that generates the motions that take us away from $\Sigma$.  On a null boundary, the situation is different. Diffeomorphisms that take us away from the null surface, e.g.\ those that are generated by vector fields $\xi^a\propto \I\uo{e}{A\bar{A}}{a}k^A\bar{k}^{\bar{A}}$, are no longer gauge symmetries. In fact, they are not even Hamiltonian, and there is a clear physical reason why this must be so.  To move the null boundary relative to the metric in the bulk, new initial data must be specified. This data is arbitrary. There are infinitely many ways to extend a given solution in the interior to the outside. Hence, such an outward radial evolution can no longer be Hamiltonian. At the quantum level, we expect, therefore, that the radial evolution will be governed by some non-unitary flow equation, e.g.\ a Lindblad equation for the physical states. In this way, the hard problem of the ADM canonical framework, namely how to solve the Wheeler--DeWitt equation, turns into a secondary question. 
First, we seek for a quantum representation of the  geometry of the null cone. As we will see below, we can already do interesting physics at that stage. 
The question of how to move the null boundary and trace out, thereby, an entire four-dimensional geometry is secondary. 
It is related to coarse graining and radial renormalization \cite{DittrichSteinhaus13,Dittrich:2014ala,Freidel:2008sh}.

\subsection{$SL(2,\R)$ holonomies on the lightcone}\label{sec2.3}
\noindent The gauge equivalence class of boundary data \eref{bndrycond} provides a complete characterization of the free radiative data on the null surface boundary. Yet, this definition is not very practical. To compute Poisson brackets among Dirac observables, we need a more useful parametrization. In the following, the basic idea is to use certain group-valued data to parametrize the boundary metric at the full non-perturbative level. Instead of the metric $q_{ab}$, we work with dyadic one-forms $(m_a,\bar{m}_a)$, which are normal to the null directions, i.e.\ $l^am_a=0$ for all $l^a\in T\mathcal{N}:\pi_\ast l^a=0$. The one-forms $m_a$ are thus characterized by four numbers. One of them can be chosen to be a conformal factor $\Omega$ that determines the overall scale of the geometry. The other three can be repackaged into an $SL(2,\R)$ holonomy that determines a transport equation between an arbitrary fiducial dyad $({}^{(0)}m_a,{}^{(0)}\bar{m}_a)$ and the physical dyads $(m_a,\bar{m}_a)$. To pick such a fiducial dyad, we choose coordinates $(z,\bar{z})$ on the base manifold $\mathcal{C}\simeq S^2$. Next, we drag the coordinates along the null generators using the Lie derivative, i.e.\ $l^a\partial_a[z]=0$ for all $l^a:\pi_\ast\ell^a=0$. We can now define the following fiducial one-forms on $\mathcal{N}$. We set
\begin{equation}
{}^{(0)}m=\frac{\sqrt{2}}{1+|z|^2}\di z.
\end{equation}
The corresponding fiducial area element is
\begin{equation}
d^2v_o=-\I\,{}^{(0)}m\wedge{}^{(0)}\bar{m}.
\end{equation}

To define an $SL(2,\R)$ group action on the dyads, we introduce an auxiliary two-dimensional real vector space $\mathbb{V}$ and its dual space $\mathbb{V}^\ast$ with complex basis vectors
\begin{align}
m^i\equiv\frac{1}{\sqrt{2}}\begin{pmatrix}1\\+\I\end{pmatrix},&\qquad\bar{m}^i\equiv\frac{1}{\sqrt{2}}\begin{pmatrix}1\\-\I\end{pmatrix},\\
m_i\equiv\frac{1}{\sqrt{2}}\begin{pmatrix}1&\I\end{pmatrix},&\qquad\bar{m}_i\equiv\frac{1}{\sqrt{2}}\begin{pmatrix}1&-\I\end{pmatrix}.
\end{align}
Next, we define the following co-dyad on $\mathcal{N}$
\begin{equation}
\ou{e}{i}{a}= m^i\bar{m}_a+\CC,
\end{equation}
where $i=1,2$ are internal $SL(2,\R)$ indices. Any such co-frame field can be now parametrized in terms of the fiducial basis by an overall conformal factor $\Omega$ and $SL(2,\R)$ holonomy $\ou{S}{i}{j}:\mathcal{N}\rightarrow SL(2,\R)$ such that
\begin{equation}
e^i = \Omega\, \ou{S}{i}{j}\big(m^j\,{}^{(0)}\bar{m}_a+\CC\big).\label{eparam}
\end{equation}
The holonomy satisfies a simple transport equation along the null generators,
\begin{equation}
\ell^a\partial_a S=\big(\varphi_{(\ell)}J+(\sigma_{(\ell)}\bar{X}+\bar{\sigma}_{(\ell)}X)\big)S,
\end{equation}
where $\sigma_{(\ell)}$ is the shear and $\varphi_{(\ell)}=-\ell^a\Gamma_a$, for the $U(1)$ connection \eref{Gammadef}. In here, we introduced a basis in $\mathfrak{sl}(2,\R)$, whose matrix elements are given by
\begin{equation}
\ou{J}{i}{k}=\I(\bar{m}^im_k-m^i\bar{m}_k)\equiv\ou{\varepsilon}{i}{k},\quad \ou{X}{i}{k}=m^im_k,\qquad \ou{\bar{X}}{i}{k}=\bar{m}^i\bar{m}_k.\label{JXXdef0}
\end{equation}
These matrices satisfy standard the commutation relations
\begin{equation}
[J,X]=-2\I\, X,\qquad[J,\bar{X}]=+2\I\, \bar{X},\qquad [X,\bar{X}]=\I\,J
\end{equation}

\subsection{Choice of relational clock}\label{sec2.4}
\noindent To access the two true physical degrees of freedom of the gravitational field on a null cone, we need to impose the Raychaudhuri equation, which is the generator on phase space of angle dependent reparametrizations of the null generators $\gamma_q=\pi^{-1}(q), q\in\mathcal{C}$. %
At the null surface, we have a fiducial, but unphysical, time coordinate $u$, which is the $2+1$ Carollian version of the usual ADM time coordinate $t$.  Surfaces of constant $u$ are two-dimensional spatial sections of $\mathcal{N}$, but they do not satisfy any particular constraint. If we now also add the coordinates $(z,\bar{z})$, which are constant along the null generators, we have a fiducial three-dimensional coordinate system $(u,z,\bar{z})$ on $\mathcal{N}$. Thus, we also have a basis $(\partial_u^a,\partial^a_z,\partial^a_{\bar{z}})$ in $T\mathcal{N}$, where $\partial_u^a$ is null. Since $k=-N\di u$, see \eref{kdef}, we obtain
\begin{equation}
\ell^a = N^{-1}\partial^a_u.\label{ell1}
\end{equation}
Dilations of  $\ell^a$, sending $\ell^a$ into $\E^{\lambda}\ell^a$ are gauge symmetries. The corresponding boost charge $K_\lambda[\mathcal{C}_\pm]$, see e.g. \cite{Wieland:2021vef}, is given by the gauge parameter $\lambda$ integrated against the area of the cross section
\begin{equation}
K_\lambda[\mathcal{C}_{\pm}]=\frac{1}{8\pi G}\oint_{\mathcal{C}_\pm}\varepsilon\,\lambda.\label{Kdef}
\end{equation}
Dilatations of the null normal change the affinity $\kappa_{(\ell)}$, sending $\kappa_{(\ell)}$ into $\E^\lambda(\kappa_{(\ell)}+\ell^a\partial_a\lambda)$.\smallskip

To proceed further, we choose a clock. Taking advantage of both dilatations and angle-dependent diffeomorphisms, we introduce a new coordinate system $(\mathcal{U},z,\bar{z})=\mathcal{U}(u,z,\bar{z}), z,\bar{z})$ on $\mathcal{N}$ in which the Raychaudhuri equation simplifies. Our choice is\footnote{We call this clock \emph{teleological}, from greek \emph{telos}, which means \emph{end}, \emph{purpose} or \emph{goal}, because our clock knows that it will assume the values $\pm 1$ at the future and past endpoints of $\mathcal{N}$. It is impossible to find such a clock from data given only in a small neighbourhood of $\mathcal{C}_-$, i.e.\ the past corner of $\mathcal{N}$.}
\begin{align}
\partial^b_{\mathcal{U}}\nabla_b\partial^a_{\mathcal{U}} & = -\Omega^{-1}\partial_{\mathcal{U}}\Omega\,\partial^a_{\mathcal{U}},\label{Ugauge}\\
\mathcal{U}\big|_{\mathcal{C}_{\pm}}& =\pm1.
\end{align}
With respect to this clock, the Raychaudhuri equation becomes
\begin{equation}
\frac{\di^2}{\di\mathcal{U}^2}\Omega^2=-2\sigma\bar{\sigma}\Omega^2.\label{Req1}
\end{equation}
A quick side remark: if there is no shear ($\sigma=0$), the area of the cross sections will increase linearly in $\mathcal{U}$ and $d=\sqrt{\mathcal{U}}$ becomes a sort-of luminosity distance. 

In here, the shear  is obtained from the transport equation for the $SL(2,\R)$ holonomy.\footnote{The spin coefficients $\varphi_{(\ell)}$ and $\sigma_{(\ell)}$ refer to the physical co-frame $(k,m,\bar{m})$, which is dual to $(\ell^a,m^a,\bar{m})$. It is this co-frame that enters the definition of the kinematical symplectic potential \eref{Theta1}. For the spin coefficients of the gauge-fixed co-frame $(-\di \mathcal{U},m,\bar{m})$, we drop the subscript and simply write $\varphi$ and $\sigma$.}
\begin{equation}
\frac{\di}{\di\mathcal{U}}S = \Big(\varphi J+\big(\sigma\bar{X}+\CC\big)\Big)S.\label{transporteq}
\end{equation}
The new time variable $\mathcal{U}$, which is a partial observable on the kinematical phase space of the null surface, can be obtained from the fiducial and unphysical time variable $u$ ($\bbvar{d}u=0$) by
\begin{equation}
\mathcal{U}(u,z,\bar{z})=-1+\int_{-1}^u\di u'N(u',z,\bar{z})\E^{\lambda(u,z,\bar{z})},\label{Udef}
\end{equation}
where $N$ is the lapse function and the gauge parameter $\lambda=\lambda^{(0)}+\Delta\lambda$ is the sum of
\begin{align}
\lambda^{(0)}(u,z,\bar{z})&=\int_{-1}^u\di u'\big(N\kappa_{(\ell)}+\Omega^{-1}\partial_u\Omega\big)(u',z,\bar{z}),\\
\E^{-(\Delta\lambda)(z,\bar{z})}&=\frac{1}{2}\int_{-1}^1\di u\, N(u,z,\bar{z})\E^{\lambda^{(0)}(u',z,\bar{z})}.
\end{align}
Thus, the clock depends as a functional on the lapse function $N$, the affinity $\kappa_{(\ell)}$ and the conformal factor $\Omega$. Each of these fields are part of the kinematical phase space equipped with the symplectic potential \eref{Theta1}. Therefore, the new clock is a field-dependent variable. The same applies to the gauge parameter $\lambda$, such that
\begin{equation}
\bbvar{d}\mathcal{U}\neq 0, \qquad \bbvar{d}\lambda\neq 0.
\end{equation}
In fact, the map between $\ell^a$ and $\partial^a_{\mathcal{U}}$ is a dilation with gauge parameter $\lambda$. Going back to \eref{ell1} and \eref{Udef} obtain
\begin{equation}
\ell^a= \E^\lambda\partial^a_{\mathcal{U}},\qquad k\wedge m\wedge\bar{m}=-\E^{-\lambda}\di\mathcal{U}\wedge{m}\wedge\bar{m}.\label{ellparam}
\end{equation}

\subsection{Null symplectic structure in terms of $SL(2,\R)$ holonomies}\label{sec2.5}
\noindent After having introduced a paramerization for the null direction $\ell^a$ and for the co-dyad $(m,\bar{m})$, see \eref{ellparam} and \eref{eparam}, we return to the pre-symplectic potential \eref{Theta1}. All differentials on phase space can be absorbed back into variations of $(\E^\lambda,\Omega,\Gamma)$ and an $SL(2,\R)$ holonomy on the lightcone. We obtain
\begin{align}
\Theta_{\mathcal{N}}=&-\frac{1}{8\pi G}\int_{\partial\mathcal{N}}d^2v_o\,\Omega^2\bbvar{d}\lambda+\frac{1}{16\pi G}\int_{\partial\mathcal{N}}d^2v_o\bbvar{d}\Omega^2+\nonumber\\
&-\frac{1}{8\pi\gamma G}\int_{\mathcal{N}}\bigg[\frac{1}{2}d^2v_o\,\wedge\di\Omega^2\,\operatorname{Tr}\big(J\bbvar{d}SS^{-1}\big)+\nonumber\\
&\hspace{6em}+d^2v_o\wedge \di\mathcal{U}\,\Omega^2\operatorname{Tr}\Big((\sigma\bar{X}+\bar{\sigma}X)(\gamma\bbvar{1}-J)\bbvar{d}SS^{-1}\Big)\bigg]+\nonumber\\
&-\frac{1}{8\pi G}\int_{\mathcal{N}}d^2v_o\,\bbvar{d}\mathcal{U}\frac{\di^2}{\di\mathcal{U}^2}\Omega^2-\frac{1}{8\pi\gamma G}\int_{\mathcal{N}}d^2v_o\,\Omega^2\bbvar{d}\Gamma,\label{Theta2}
\end{align}
where  $\Gamma=-\varphi\,\di \mathcal{U}+\gamma\bar{m}+\bar{\gamma}m$ is the $U(1)$ connection \eref{Gammadef} on $\mathcal{N}$ and $(J,X,\bar{X})$ are the $SL(2,\R)$ generators \eref{JXXdef0}.
In what follows, the two boundary terms in the first line are irrelevant: The second term is a total derivative on phase space. Hence, it does not contribute to the presymplectic two-form. The first term, on the other hand, can be reabsorbed back into a redefinition of the $U(1)$ connection.\footnote{This is possible, because the generator $K_\lambda[\mathcal{C}]$ for dilations that acts on the triple $(\ell^a,m_a,\bar{m}_a)$ by $\delta_\lambda^{\mtext{boost}}(\ell^a,m_a,\bar{m}_a)=(\lambda\ell^a,m_a,\bar{m}_a)$ and the generator $L_\lambda[\mathcal{C}]$ of $U(1)$ frame rotations that acts by $\delta_\lambda^{rot}(\ell^a,m_a,\bar{m}_a)=(\ell^a,\I\lambda m_a,-\I\lambda\bar{m}_a)$ are proportional, i.e.\ $K_\lambda[\mathcal{C}]=\gamma L_\lambda[\mathcal{C}]$. This constraint plays an important role in covariant approaches to loop quantum gravity \cite{liftng,LQGvertexfinite}.} We can thus set them both to zero.\smallskip

Next, we need to further simplify this expression and remove the residual gauge reparametrization symmetry. To this goal, we introduce a type of interaction picture, where we seperate the $U(1)$ part of the transport equation \eref{transporteq} from the shear degrees of freedom. We set
\begin{equation}
S=RHS_-= RS_I,\label{Sinteract}
\end{equation}
where $R$ is the $U(1)$ holonomy along the light rays that generate the null surface. We have
\begin{equation}
R=\E^{\Delta J},\quad \Delta(\mathcal{U},z,\bar{z})=\int_{-1}^{\mathcal{U}}\di\mathcal{U}'\,\varphi(\mathcal{U}',z,\bar{z}),
\end{equation}
where $J$ is the complex structure introduced in \eref{JXXdef0}. Using the interaction picture, the new $SL(2,R)$ holonomy $H$ satisfies the transport equation 
\begin{equation}
\frac{\di}{\di\mathcal{U}}H=(\sigma_I\bar{X}+\bar{\sigma}_IX)H,\label{Holdef}
\end{equation}
for the shear in the interaction picture
\begin{equation}
\sigma_I(\mathcal{U},z,\bar{z})=\sigma_{(\ell)}(\mathcal{U},z,\bar{z})\E^{-2\I \Delta(\mathcal{U},z,\bar{z})}.\label{sigmaintpic}
\end{equation}
The initial conditions are
\begin{align}
H(\mathcal{U}=-1,z,\bar{z})&=\bbvar{1},\\
S_I(\mathcal{U}=-1,z,\bar{z})&=S_-(z,\bar{z}).
\end{align}
As a shorthand notation, we will also write $S\big|_{\mathcal{C}_\pm}=S_\pm$ and
\begin{align}
S_+(z,\bar{z})&=R_+H(\mathcal{U}=+1)S_-\equiv\E^{J\Delta_+}H(\mathcal{U}=+1)S_-,\label{Sfinal}\\
\Delta_+(z,\bar{z})&=\Delta(\mathcal{U}=+1,z,\bar{z}).
\end{align}

To simplify \eref{Theta2} and impose the Raychaudhuri equation, we perform a partial integration in the third term of equation \eref{Theta2}. In this way, we obtain an exterior derivative that hits the field space Maurer--Cartan form $\bbvar{d}SS^{-1}$. The resulting expression satisfies
\begin{align}
d^2v_o\wedge\di\left(\bbvar{d}SS^{-1}\right)&= d^2v_o\wedge S\bbvar{d}\left(S^{-1}\di S\right)S^{-1}=\nonumber\\
&=-d^2v_o\wedge\bbvar{d}\Gamma\,J+d^2v_o\wedge R\,\bbvar{d}\big(\di\mathcal{U}\bar{\sigma}_I+\di\mathcal{U}\sigma_I\bar{X}\big)R^{-1}+\nonumber\\
&\quad+d^2v_o\wedge\Gamma\,R\left[\bbvar{d}S_IS_I^{-1},J\right]R^{-1}+\nonumber\\
&\quad-d^2v_o\wedge R\left[\bbvar{d}S_IS_I^{-1},\di S_IS_I^{-1}\right]R^{-1}.
\end{align}
Next, we insert this eqaution back into the pre-symplectic structure. We obtain
\begin{align}
\Theta_{\mathcal{N}}&=-\frac{1}{16\pi\gamma G}\int_{\partial{\mathcal{N}}}d^2v_o\,\Omega^2\,\operatorname{Tr}\big(J\bbvar{d}SS^{-1}\big)+\nonumber\\
&\quad-\frac{1}{8\pi G}\int_{\mathcal{N}}\di\mathcal{U}\wedge d^2v_o\Omega^2\,\operatorname{Tr}\big((\sigma_I\bar{X}+\bar{\sigma}_IX)\bbvar{D}S_IS_I^{-1}\big)\nonumber+\\
&\quad-\frac{1}{ 8\pi G}\int_{\mathcal{N}}\di\mathcal{U}\wedge d^2v_o\,\bbvar{d}\mathcal{U}\Big(\frac{\di^2}{\di\mathcal{U}^2}\Omega^2+2\sigma_I\bar{\sigma}_I\Omega^2\Big).\label{Theta3}
\end{align}
The last term vanishes on-shell, i.e.\ if the Raychaudhuri constraint \eref{Req1} is satisfied. It also shows, not very surprisingly, that the clock time $\mathcal{U}$ is conjugate to the constraint. Furthermore, we introduced a field space covariant derivative $\bbvar{D}$, which differs from the usual exterior derivative on phase space by a vertical diffeomorphism such that it annihilates the clock variable. In other words, we now have that
\begin{align}
\bbvar{D}S_I&=\bbvar{d}S_I-\bbvar{d}\mathcal{U}\big(\sigma_I\bar{X}+\bar{\sigma}X\big)S_I=\bbvar{d}S_I-\bbvar{d}\mathcal{U}\frac{\di}{\di\mathcal{U}}S_I,\\
\bbvar{D}\Omega&=\bbvar{d}\Omega-\bbvar{d}\mathcal{U}\frac{\di}{\di\mathcal{U}}\Omega,\\
\bbvar{D}\mathcal{U}&=0.
\end{align}

Equation \eref{Theta3} is a local version of the standard symplectic structure of $T^\ast SL(2,\R)$. It is interesting to note that, by adding the Barbero--Immirzi parameter to the action, new entries appear in the symplectic structure that would  vanish as $\gamma\rightarrow \infty$. In this way,  representations of $T^\ast SL(2,\C)$ can now suddenly count that do otherwise not. It is this mechanism that is responsible for the discreteness of geometry in loop quantum gravity. 

\subsection{Variation of the $SL(2,\R)$ holonomy}\label{sec2.6}
\noindent To compute the Poisson brackets among Dirac observables, we will crucially need the functional derivative of the holonomy \eref{Holdef}. Since, however, the covariant field space derivative $\bbvar{D}$ and the time derivative $\frac{\di}{\di\mathcal{U}}$ commute, the calculation is straightforward. Going back to the defining differential equation for the holonomy, i.e.\ \eref{Holdef} and taking a functional derivative, we obtain
\begin{equation}
H^{-1}\frac{\di}{\di\mathcal{U}}\bbvar{D}H=H^{-1}\left(\bbvar{D}\sigma_{I}+\bbvar{D}\bar{\sigma}_IX\right)H-\frac{\di}{\di\mathcal{U}}\left(H^{-1}\right)\bbvar{D}H.
\end{equation}
Next, we multiply this equation from the left by $H(\mathcal{U})$ and integrate the resulting expression from $\mathcal{U}'=-1$ to $\mathcal{U}$. Taking into account that $\bbvar{D}H(\mathcal{U}=-1)=\bbvar{D}\bbvar{1}=0$, we see
\begin{equation}
(\bbvar{D}H)(\mathcal{U})=\int_{-1}^{\mathcal{U}}\di\mathcal{U}'H(\mathcal{U})\left(H^{-1}(\bbvar{D}\sigma_I\bar{X}+\bbvar{D}\bar{\sigma}_IX)H\right)(\mathcal{U'}).\label{Holvar}
\end{equation}

\section[Twistor quantisation of the unimodular group]{Twistor quantisation of ${T^\ast SL(2,\mathbb{R})}$}\label{sec3}
\subsection{Twistor parametrisation of $T^\ast SL(2,\R)$}\label{sec3.1}
\noindent To quantise the geometry of the light cone, we  now turn to the phase space $T^\ast SL(2,\R)$ and its quantisation. Since $T^\ast SL(2,\R)$ is isomorphic to $T^\ast SU(1,1)$ and since it is better to work with complex numbers rather than with real numbers, we can work with $T^\ast SU(1,1)$ instead.  The standard symplectic structure of $T^\ast SU(1,1)$ is
\begin{equation}
\Theta_{SU(1,1)} = \operatorname{Tr}(\Pi\di UU^{-1}), \quad\text{where}\quad (\Pi,U)\in\mathfrak{su}(1,1)\times SU(1,1).
\end{equation}
To split the momentum into its rotational and translational components, we introduce the following matrices, 
\begin{equation}
J=\begin{pmatrix}
\I&0\\
0&-\I
\end{pmatrix},\quad
X=\begin{pmatrix}
0&0\\
1&0
\end{pmatrix},\quad
\bar{X}=\begin{pmatrix}
0&1\\
0&0
\end{pmatrix}.\label{JXXdef1}
\end{equation}
Notice that all products and traces are isomorphic to \eref{JXXdef0}. Angular and translational components $(L,c,\bar{c})$ of $\Pi$ are now given by
\begin{equation}
\Pi = LJ+(c\bar{X}+\bar{c}X).
\end{equation}
The group $SU(1,1)$ has a standard action on $\C^2$ that preserves the signature $(+,-)$ inner product between spinors $U^A,V^A\in\C^2$:
\begin{equation}
\eta_{A\bar{A}}U^A\bar{V}^{\bar{A}}=U^0\overline{{V}^{0}}-U^1\overline{{V}^{1}}.
\end{equation}
In addition, the group is unimodular. For any $\mathfrak{sl}(2,\R)$ element $\ou{\Pi}{A}{B}$, we thus find the two conditions
\begin{align}
\Pi_{AB}&=\Pi_{BA},\\
\Pi_{AB}&=-\eta_{A\bar{A}}\eta_{B\bar{B}}\bar{\Pi}^{\bar{A}\bar{B}},\label{hcty}
\end{align}
where spinor indices are raised and lowered using the skew symmetric $\epsilon$-spinor, e.g.\ $\Pi_{AB}=\epsilon_{CA}\ou{\Pi}{C}{B}$, $\mathrm{Tr}(\Pi)=\ou{\Pi}{A}{A}=\epsilon^{BA}\Pi_{AB}=0$.\smallskip

Any symmetric spinor of rank $n$ can be split into the symmetrised tensor products of $n$ eigenspinors \cite{penroserindler}. Therefore, we can always find spinors $\pi_A$ and $\omega_A$ such that,
\begin{equation}
\Pi_{AB}=\pi_{(A}\omega_{B)}.
\end{equation}
In terms of these spinors, the $SL(2,\R)$ Casimir is simply given by
\begin{equation}
\mathrm{Tr}(\Pi^2)=-2\left(L^2-c\bar{c}\right)=-\Pi_{AB}\Pi^{AB}=\frac{1}{2}(\pi_A\omega^A)^2
\end{equation}

The spinors are unique up to the discrete exchange of $\pi_A$ and $\omega_A$ and complexified dilations mapping $\pi_A\rightarrow\E^z\pi_A$ and $\omega_A\rightarrow\E^{-z}\omega_A$ for some $z\in\C$. To obtain an $\mathfrak{su}(1,1)$ Lie algebra element, we now also have to satisfy \eref{hcty}. There are two cases to distinguish that correspond to the discrete and continuous series representations of $SL(2,\R)$. In the first case, we have
\begin{equation}
\exists z\in\C:\left(\pi^A,\omega^A\right)=\left(\E^z\ou{\eta}{A}{\bar{A}}\bar{\pi}^{\bar{A}},-\E^{-z}\ou{\eta}{A}{\bar{A}}\bar{\omega}^{\bar{A}}\right).
\end{equation}
We can then always rescale the spinors $(\pi^A,\omega^A)$ such that we have the simplified condition 
\begin{equation}
\left(\pi^A,\omega^A\right)=\left(\ou{\eta}{A}{\bar{A}}\bar{\pi}^{\bar{A}},-\ou{\eta}{A}{\bar{A}}\bar{\omega}^{\bar{A}}\right).\label{case1}
\end{equation}
The first case corresponds, therefore, to a configuration in which
\begin{equation}
\pi_A\omega^A=\bar{\pi}_{\bar{A}}\bar{\omega}^{\bar{A}}\Leftrightarrow c\bar{c}\geq L^2.
\end{equation}
In the second case, on the other hand, we have 
\begin{equation}
\exists z\in\C:\left(\pi^A,\omega^A\right)=\left(\E^z\ou{\eta}{A}{\bar{A}}\bar{\omega}^{\bar{A}},-\E^{-z}\ou{\eta}{A}{\bar{A}}\bar{\pi}^{\bar{A}}\right).\label{case2}
\end{equation}
Again, we can reabsorb $z$ back into the definition of the two eigenspinors of $\Pi_{AB}$. We are then only left with two possibilities,
\begin{equation}
{}^\pm\pi^A=\pm\I\ou{\eta}{A}{\bar{A}}{}^\pm\bar{\omega}^{\bar{A}}
\end{equation}
Hence, there are two series, which we distinguish in our notation by writing $({}^\pm\pi^A,{}^\pm\omega^A)$ for either sign. The second case corresponds, therefore, to a configuration in which
\begin{equation}
\pi_A\omega^A=-\bar{\pi}_{\bar{A}}\bar{\omega}^{\bar{A}}\Leftrightarrow c\bar{c}\leq L^2.
\end{equation}

In the following, we exclude the case in which the Casimir vanishes, such that $\pi_A\omega^A\neq0$. If this condition is satisfied, the pair $(\pi^A,\omega^A)$ forms a basis in $\C^2$. This implies that the configuration variable $\ou{U}{A}{B}\in SL(2,\R)$ is completly characterized by how it acts on this basis. We thus define a new pair $(\utilde{\pi}^A,\utilde{\omega}^A)$ of spinors such that
\begin{equation}
\omega^A=\ou{U}{A}{B}\utilde{\omega}^B,\quad\pi^A=\ou{U}{A}{B}\utilde{\pi}^B.\label{leftspin}
\end{equation}
Since $\ou{U}{A}{B}$ has determinant one, these spinors satisfy the matching condition
\begin{equation}
C=\pi_A\omega^A-\utilde{\pi}_A\tilde{\omega}^A=0.\label{matchcond}
\end{equation}

Given the quadruple $(\pi^A,\omega^A,\utilde{\pi}^A,\utilde{\omega}^A)$, we can reconstruct both $U\in SU(1,1)$ and $\Pi\in\mathfrak{su}(1,1)$. We can thus take this parametrisation and insert it back into the symplectic potential. A short calculation gives
\begin{align}
\Theta_{SU(1,1)} = \operatorname{Tr}(\Pi\di UU^{-1})&=\pi_{(A}\omega_{B)}\di\ou{U}{A}{C}U^{BC}=\pi_A\omega_B\di\ou{U}{A}{C}U^{BC}=\pi_A\di\ou{U}{A}{C}\utilde{\omega}^C=\nonumber\\
&=\pi_A\di(\ou{U}{A}{C}\utilde{\omega}^C)-\pi_A\ou{U}{A}{C}\di\utilde{\omega}^C=\pi_A\di\omega^A-\utilde{\pi}_A\di\utilde{\omega}^A.\label{ThetaSL2R}
\end{align}
For either case $\pi_A\omega^A\gtrless0$, we can thus embed $\Theta_{SU(1,1)} $ into a larger phase, which is now formed by the spinors $(\pi^A,\omega^A,\utilde{\pi}^A,\utilde{\omega}^A)$ subject to constraints \eref{matchcond} and \eref{case1} and \eref{case2}. The entire construction is in complete analogy with the spinor and twistor representation of quantum geometry \cite{Girelli:2005ii,Dupuis:2014fya,twist,Bianchi:2016tmw,Bianchi:2016hmk,Livinerep,spezialetwist1,twistintegrals,komplexspinors}.

\subsection{Twistor quantisation of $T^\ast SL(2,\R)$: continuous series representations}\label{sec3.2}
\noindent There are two cases to consider. In the first case, $c\bar{c}\geq L^2$. In this case, the spinors satisfy the reality condition \eref{case1}. The solution of this equation is immediate to find. There are phase space variables $(p,q)\in\C^2$ such that
\begin{equation}
\begin{pmatrix}
\pi^0\\\pi^1
\end{pmatrix}=
\begin{pmatrix}
p\\-\bar{p}
\end{pmatrix},\quad
\begin{pmatrix}
\omega^0\\\omega^1
\end{pmatrix}=
\begin{pmatrix}
\bar{q}\\q
\end{pmatrix},
\end{equation}
and equally for $(\utilde{\pi}^A,\utilde{\omega}^A)$ in terms of $(\utilde{p},\utilde{q})$. Going back to the symplectic potential \eref{ThetaSL2R}, we infer the canonical commutatiuon relations
\begin{equation}
\{p,q\}=1,\qquad\{\bar{p},\bar{q}\}=1.
\end{equation}
Going to a position representation, we obtain wavefunctions
\begin{equation}
\langle q,\bar{q}|\lambda,m\rangle = \frac{1}{2\pi}(q\bar{q})^{\frac{1}{2}(\I\lambda-1)}\left(\frac{q}{\bar{q}}\right)^{m},
\end{equation}
where $\lambda\in\R$ and $m\in\Z/2$. The normalisation is $\langle\lambda,m|\lambda',m'\rangle=\delta(\lambda-\lambda')\delta_{mm'}$.
On the corresponding Hilbert space $L^2(\C,\frac{\I}{2}\di q\,\di\bar{q})$, the $SL(2,\R)$ generators admit the representation
\begin{align}
L&=-\frac{1}{2}\mathrm{Tr}(PJ)=\frac{1}{2}\left(q\partial_q-\bar{q}\partial_{\bar{q}}\right),\\
c&=\mathrm{Tr}(PX)=-\I\bar{q}\partial_q,\label{cqdef1}\\
c^\dagger\equiv\bar{c}&=\mathrm{Tr}(P\bar{X})=-\I{q}\partial_{\bar{q}},
\end{align}
The operators act as
\begin{align}
L|\lambda,m\rangle&=m|\lambda,m\rangle,\\
c|\lambda,m\rangle&=-\frac{\I}{2}\left(\I\lambda+2m-1\right)|\lambda,m-1\rangle,\\
\bar{c}|\lambda,m\rangle&=-\frac{\I}{2}\left(\I\lambda-2m-1\right)|\lambda,m+1\rangle,\\
\left(L^2-\tfrac{1}{2}(c\bar{c}+\bar{c}c)\right)|\lambda,m\rangle&=-\frac{1}{4}(\lambda^2+1)|\lambda,m\rangle.
\end{align}

\subsection{Twistor quantisation of $T^\ast SL(2,\R)$: discrete series representations}\label{sec3.3}
\noindent The second case happens when the $SL(2,\R)$ Casimir satisfies $L^2>c\bar{c}$. In this case, the symplectic structure \eref{ThetaSL2R} implies that the coefficients of the $SL(2,\R)$ spinors are canonical creation and annihilation operators
\begin{alignat}{3}
\begin{pmatrix}
{}^+\omega^0\\
{}^+\omega^1
\end{pmatrix}
&=
\begin{pmatrix}
d\\
e^\dagger
\end{pmatrix}
,&\qquad 
\begin{pmatrix}
{}^-\omega^0\\
{}^-\omega^1
\end{pmatrix}
&=
\begin{pmatrix}
d^\dagger\\
e
\end{pmatrix},\\
\begin{pmatrix}
{}^+\pi^0\\
{}^+\pi^1
\end{pmatrix}
&=
-\I\begin{pmatrix}
e\\
d^\dagger
\end{pmatrix}
,& 
\begin{pmatrix}
{}^-\pi^0\\
{}^-\pi^1
\end{pmatrix}
&=\I
\begin{pmatrix}
e^\dagger\\
d
\end{pmatrix}.
\end{alignat}
The fundamental canonical commutation relations are
\begin{equation}
[d,d^\dagger]=[e,e^\dagger]=1.
\end{equation}
All other commutators vanish. The resulting representation of the $SL(2,R)$ generators is
\begin{align}
{}^\pm L&=\pm\frac{1}{2}\left(d^\dagger d+e^\dagger e+1\right),\\
{}^+c&={}^-c^\dagger=-\I de,\label{cqdef2}\\
L^2-\tfrac{1}{2}(cc^\dagger+c^\dagger c)&=\frac{1}{4}\left(d^\dagger d-e^\dagger e-1\right)\left(d^\dagger d-e^\dagger e+1\right).
\end{align}

\section{Quantum impulsive data on the null cone\label{sec4}}
\subsection{Phase space of a single pulse of radiative data}\label{sec4.1}
\noindent In this section, we carry out the main task of this paper. The idea is to start from a truncation of the classical phase space, in which we only consider a specific class of impulsive null initial data. By restricting ourselves to a finite such puls, the infinite-dimensional gravitational bulk and boundary phase space is reduced to a finite-dimensional mechanical system with constraints. This truncation provides a great simplification. At the end of the section, we can take the analysis to the quantum level and solve the constraints at the full non-perturbative level.\smallskip

The first step is to restrict the phase space equipped with the null symplectic potential \eref{Theta3} to a single pulse. A large but finite sequences of such pulses, corresponding to a sequence of null slabs $(\mathcal{N}_1,\mathcal{N}_2,\dots)$ can approximate to arbitrarily good precision any smooth shear $\sigma_{(\ell)}$ with compact support on $\mathcal{K}=\mathcal{N}_1\cup\mathcal{N}_2\cup\dots$.\footnote{As a Cauchy sequence with respect to an auxiliary inner product on $\mathcal{K}$, e.g.\ $L^2(\mathcal{K},d^2v_o\wedge k)$.} At the quantum level, we will then have a corresponding sequence of Hilbert spaces $(\mathcal{H}_{\mathcal{N}_1},\mathcal{H}_{\mathcal{N}_2},\dots)$. The physical Hilbert space for the entire boundary data will be then built from the tensor product $\mathcal{H}_{\mtext{kin}}=\mathcal{H}_{\mathcal{N}_1}\otimes\mathcal{H}_{\mathcal{N}_2}\otimes\dots$ by tracing over intermediate edge states at the two-dimensional corners $\mathcal{C}_i=\partial\mathcal{N}_i\cap\partial\mathcal{N}_{i+1}$.\smallskip

In each pulse, we shall restrict ourselves to configurations with constant shear. This creates a conceptual problem. The shear $\sigma_{(\ell)}$ of a null generator $\ell^a$ of $\mathcal{N}$ is a component of the spin connection, which is a one-form in the bulk. A component of a connection does not define a scalar. Thus, our statement can only be true in a very specific gauge. At this point, we are free to choose whatever gauge we like, in particular a gauge in which the Raychaudhuri equation, which is the generator for angle dependent time reparametrizations $\mathcal{U}\rightarrow \mathcal{U}'(\mathcal{U},z,\bar{z})$, $\mathcal{U}'(\pm1,z,\bar{z})=\pm 1$, simplifies. In addition, there are further gauge symmetries we can use. There is a $U(1)$ gauge freedom and the freedom to rescale $\mathcal{\ell}^a$. All three gauge transformations are required to define the shear in the interaction picture, as we saw in \eref{sigmaintpic} above. It is the shear in this gauge that we restrict to be constant along the null generators. Our requirement can be thus summarized as
\begin{equation}
\frac{\di}{\di\mathcal{U}}\sigma_I=0,\qquad\sigma_I(\mathcal{U},z,\bar{z})=:\sigma_I(z,\bar{z}).\label{shearpulse}
\end{equation}

For constant such shear, we can immediately integrate the Raychaudhuri equation \eref{Req1}. We obtain
\begin{equation}
\Omega^2=(E_++E_-)\frac{\sin(\sqrt{2\sigma_I\bar{\sigma}_I})}{\sin(2\sqrt{2\sigma_I\bar{\sigma}_I})}\cos(\sqrt{2\sigma_I\bar{\sigma}_I}\mathcal{U})+
(E_+-E_-)\frac{\cos(\sqrt{2\sigma_I\bar{\sigma}_I})}{\sin(2\sqrt{2\sigma_I\bar{\sigma}_I})}\sin(\sqrt{2\sigma_I\bar{\sigma}_I}\mathcal{U}),\label{OmegaUdep}
\end{equation}
where $E_\pm$ are the initial conditions
\begin{equation}
E_\pm(z,\bar{z})=\Omega^2(\mathcal{U}=\pm1,z,\bar{z}).
\end{equation}
At this point, we may wish to restrict the norm of the shear to a finite interval $0\leq |\sigma_I|<\frac{\pi}{2\sqrt{2}}$. However, no such restriction seems necessary in our final result. In any case, at null infinity, we are well within this regime. During its finite duration, the pulse will have the asymptotic form $\sigma_I(z,\bar{z})=\frac{\dot{\sigma}^{(0)}(z,\bar{z})}{r}+\mathcal{O}(r^{-2})$. See also \hyperref[shearnote]{\color{darkblue} footnote 11} for a similar remark, which will be made more precise elsewhere.\smallskip

We then also have two integrals to solve. They are
\begin{align}
\int_{-1}^1\di\mathcal{U}\,\Omega^2(\mathcal{U},z,\bar{z})&=+(E_++E_-)\frac{\tan(\sqrt{2\sigma_I\bar{\sigma}_I})}{\sqrt{2\sigma_I\bar{\sigma}_I}},\label{int1}\\
\int_{-1}^1\di\mathcal{U}\,\mathcal{U}\,\Omega^2(\mathcal{U},z,\bar{z})&=-(E_+-E_-)\frac{1}{\sin(\sqrt{2\sigma_I\bar{\sigma}_I})}\frac{\di}{\di \rho}\left[\frac{\sin(\rho)}{\rho}\right]\bigg|_{\rho=\sqrt{2\sigma_I\bar{\sigma}_I}}\label{int2}.
\end{align}
These integrals will prove useful when evaluating the pre-symplectic potential on-shell, i.e.\ when the Raychaudhuri equation is satisfied.\smallskip

Next, we need to speak about the $SL(2,\R)$ holonomy \eref{Holdef}. Since the shear is now constant along the null generators, we can forget about the time-ordering, and immediately integrate the defining differential equation \eref{Holdef}. We obtain,
\begin{equation}
H=\ch\left(\sqrt{\sigma_I\bar{\sigma}_I}(\mathcal{U}+1)\right)\bbvar{1}+\frac{1}{\sqrt{\sigma_I\bar{\sigma}_I}}\left(\bar{\sigma}_IX+{\sigma}_I\bar{X}\right)\sh\left(\sqrt{\sigma_I\bar{\sigma}_I}(\mathcal{U}+1)\right),\label{HolUdep}
\end{equation}
where we again supressed all angular dependence, i.e.\ $H\equiv H(\mathcal{U},z,\bar{z})$.\smallskip

We can now turn to our main task in this section and compute the physical phase space of the system. The strategy is simple. We insert the solutions for the $SL(2,\R)$ holonomy $H(\mathcal{U},z,\bar{z})$ and the conformal factor $\Omega^2(\mathcal{U},z,\bar{z})$, see \eref{HolUdep} and \eref{OmegaUdep}, back into the pre-symplectic structure \eref{Theta3}, thereby calculating the pull-back from the kinematical phase space to the physical phase space of the radiative pulse. In doing so, we have to take into account also the boundary conditions \eref{Sinteract} and \eref{Sfinal} in the interaction picture.\smallskip 

In what follows, we split the calculation into two steps. First of all, we consider the term that contains the bulk integral along $\mathcal{N}$. Taking into account the variation of the holonomy \eref{Holvar}, we obtain
\begin{align}
&\int_{\mathcal{N}}d^2v_o\wedge\di\mathcal{U}\,\Omega^2\operatorname{Tr}\left((\sigma_I\bar{X}+\bar{\sigma}_IX)\bbvar{D}S_IS_I^{-1}\right)=\\
&=\int_{-1}^1\di\mathcal{U}\int_{-1}^{\mathcal{U}}\di\mathcal{U}'\int_{\mathcal{C}}d^2v_o\,\Omega^2(\mathcal{U})\operatorname{Tr}\left((\sigma_I\bar{X}+\CC)H(\mathcal{U},\mathcal{U}')\left(\bbvar{D}\sigma_I\bar{X}+\CC\right)H^{-1}(\mathcal{U},\mathcal{U}')\right)+\nonumber\\
&\quad+\int_{\mathcal{N}}d^2v_o\wedge\di\mathcal{U}\,\Omega^2(\mathcal{U})\operatorname{Tr}\left((\sigma_I\bar{X}+\CC)H(\mathcal{U})(\bbvar{D}S_-S_-^{-1})H^{-1}(\mathcal{U})\right),
\end{align}
where $H(\mathcal{U},\mathcal{U}'):=H(\mathcal{U})H^{-1}(\mathcal{U}')$ and all variables have an implicit angular dependence dropped from our notation for brevity, e.g.\ $\Omega^2(\mathcal{U})\equiv\Omega^2(\mathcal{U},z,\bar{z})$. Since, however, the shear is assumed to be constant for the entire duration of the pulse, see \eref{shearpulse}, the holonomies commute with the connection, i.e.\ $[H(\mathcal{U}),\sigma_I\bar{X}+\CC]=0$. This simplifies the integrals a lot. In fact,
\begin{align}
&\int_{\mathcal{N}}d^2v_o\wedge\di\mathcal{U}\,\Omega^2\operatorname{Tr}\left((\sigma_I\bar{X}+\bar{\sigma}_IX)\bbvar{D}S_IS_I^{-1}\right)
=\int_{-1}^1\di\mathcal{U}\int_{\mathcal{C}}d^2v_o\,\Omega^2(\mathcal{U})(1+\mathcal{U})\bbvar{D}(\sigma_I\bar{\sigma}_I)+\nonumber\\
&\quad+\int_{\mathcal{N}}d^2v_o\wedge\di\mathcal{U}\,\Omega^2(\mathcal{U})\operatorname{Tr}\left((\sigma_I\bar{X}+\CC)(\bbvar{D}S_-S_-^{-1})\right)=\nonumber\\
&=-\frac{1}{2}\int_{\mathcal{C}}d^2v_o\,(E_+-E_-)\frac{1}{\sin(\sqrt{2\sigma_I\bar{\sigma}_I)}}\frac{\di}{\di\rho}\left[\frac{\sin\rho}{\rho}\right]\bigg|_{\rho=\sqrt{2\sigma_I\bar{\sigma}_I}}\bbvar{D}(2\sigma_I\bar{\sigma}_I)+\nonumber\\
&\quad+\frac{1}{2}\int_{\mathcal{C}}d^2v_o\,(E_++E_-)\frac{\tan(\sqrt{2\sigma_I\bar{\sigma}_I})}{\sqrt{2\sigma_I\bar{\sigma}_I)}}\bbvar{D}(2\sigma_I\bar{\sigma}_I)+\nonumber\\
&\quad+\int_{\mathcal{C}}d^2v_o\,(E_++E_-)\frac{\tan(\sqrt{2\sigma_I\bar{\sigma}_I})}{\sqrt{2\sigma_I\bar{\sigma}_I}}\operatorname{Tr}\left((\sigma_I\bar{X}+\CC)(\bbvar{D}S_-S_-^{-1})\right)=\nonumber\\
&=-\int_{\mathcal{C}}d^2v_o\,(E_+-E_-)\bbvar{D}\left[\ln\left(\frac{\sin(\sqrt{2\sigma_I\bar{\sigma}_I})}{\sqrt{2\sigma_I\bar{\sigma}_I}}\right)\right]+\nonumber\\
&\quad-\int_{\mathcal{C}}d^2v_o\,(E_++E_-)\bbvar{D}\left[\ln\left(\cos(\sqrt{2\sigma_I\bar{\sigma}_I})\right)\right]+\nonumber\\
&\quad+\int_{\mathcal{C}}d^2v_o\,(E_++E_-)\frac{\tan(\sqrt{2\sigma_I\bar{\sigma}_I})}{\sqrt{2\sigma_I\bar{\sigma}_I}}\operatorname{Tr}\left((\sigma_I\bar{X}+\CC)(\bbvar{D}S_-S_-^{-1})\right).
\label{step1}
\end{align}

In our next step, we consider the boundary term that appears in the first line of \eref{Theta3}. Since $S_+=R(\mathcal{U}=1)H(\mathcal{U}=1)S_-$, see \eref{Sfinal}, we have
\begin{align}
\operatorname{Tr}\left(J\bbvar{D}S_+S_+^{-1}\right)&=-2\bbvar{D}\Delta_++\operatorname{Tr}\left(J(\bbvar{D}H)(\mathcal{U}=+1)H^{-1}(\mathcal{U}=+1)\right)+\nonumber\\
&\quad+\operatorname{Tr}\left(H^{-1}(\mathcal{U}=+1)JH(\mathcal{U}=+1)\bbvar{D}S_-S_-^{-1}\right).\label{step2}
\end{align}
To calculate the second term in this equation, we recall that $H(\mathcal{U})=\exp((1+\mathcal{U})(\sigma_I\bar{X}+\CC))$. Since the complex structure $J$ acts on the translational $SL(2,\R)$ generators by $JX=-XJ$ and $J\bar{X}=-\bar{X}J$, see \eref{JXXdef1}, we have $JH(\mathcal{U})=H^{-1}(\mathcal{U})J$. Going back to \eref{HolUdep}, we obtain
 \begin{align}
&\operatorname{Tr}\left(H^{-1}(\mathcal{U}=1)JH(\mathcal{U}=1)\bbvar{D}S_-S_-^{-1}\right)=\operatorname{Tr}\left(H^{-1}(\mathcal{U}=3)J\bbvar{D}S_-S_-^{-1}\right)=\nonumber\\
&=\ch\left(4\sqrt{\sigma_I\bar{\sigma}_{I}}\right)\operatorname{Tr}\left(J\bbvar{D}S_-S_-^{-1}\right)-\I\frac{\sh(4\sqrt{\sigma_I\bar{\sigma}_I}}{\sqrt{\sigma_I\bar{\sigma}_I}}\operatorname{Tr}\left((\sigma_I\bar{X}-\bar{\sigma}_IX)\bbvar{D}S_-S_-^{-1}\right).\label{step2.1}
\end{align}
Next, we also need the Maurer--Cartan form $\bbvar{D}HH^{-1}$, which can be inferred from \eref{HolUdep}. A short calculation gives
\begin{align}
(\bbvar{D}H)H^{-1}\big|_{\mathcal{U}=+1}
&=\frac{\I}{2}\frac{1}{\sqrt{\sigma_I\bar{\sigma}_I}}\left(\sigma_I\bar{X}-\bar{\sigma}_IX\right)\sh(4\sqrt{\sigma_I\bar{\sigma}_I})\bbvar{D}\phi_I+\nonumber\\
&\quad-J\sh^2(2\sqrt{\sigma_I\bar{\sigma}_I})\bbvar{D}\phi_I+\frac{2}{\sqrt{\sigma_I\bar{\sigma}_I}}\left(\sigma_I\bar{X}+\bar{\sigma}_IX\right)\bbvar{D}(\sigma_I\bar{\sigma}_I),
\label{step2.2}
\end{align}
where $\phi_I\equiv\phi_I(z,\bar{z})$ denotes the $U(1)$ angle
\begin{equation}
\sigma_I=|\sigma_I|\E^{\I\phi_I}.
\end{equation}

Finally, we insert the results of the first step \eref{step1}, and the second step, which are collected in \eref{step2}, \eref{step2.1}, \eref{step2.2}, back into the pre-symplectic potential, which turns, thereby, into the symplectic potential for a single pulse of radiation,
\begin{align}
\Theta^{\mtext{pulse}}_{\mathcal{N}}&=\frac{1}{8\pi\gamma G}\int_{\mathcal{C}}d^2v_o E_+\bbvar{D}\Delta_++\frac{1}{16\pi\gamma G}\int_{\mathcal{C}}d^2v_oE_+\left(1-\ch(4\sqrt{\sigma_I\bar{\sigma}_I})\right)\bbvar{D}\phi_I+\nonumber\\
&\quad+\frac{1}{8\pi G}\int_{\mathcal{C}}d^2v_o(E_+-E_-)\bbvar{D}\left[\ln\left(\frac{\sin(\sqrt{2\sigma_I\bar{\sigma}_I})}{\sqrt{2\sigma_I\bar{\sigma}_I}}\right)\right]+\nonumber\\
&\quad+\frac{1}{8\pi G}\int_{\mathcal{C}}d^2v_o(E_++E_-)\bbvar{D}\left[\ln\left(\cos(\sqrt{2\sigma_I\bar{\sigma}_I})\right)\right]+
\int_{\mathcal{C}}d^2v_o\mathrm{Tr}\left(\Pi'\bbvar{D}S_-S_-^{-1}\right),\label{ThetaP1}
\end{align}
where we introduced the $SL(2,\R)$ momentum variable\footnote{The introduction of primed variables should be clear in a moment.}
\begin{equation}
\Pi'=J L+ c'\bar{X}+\bar{c}^\prime X.
\end{equation}
The $SL(2,\R)$ momentum splits into the following generators on phase space
\begin{align}
L&=\frac{1}{16\pi\gamma G}\left(E_--\ch(4\sqrt{\sigma_I\bar{\sigma}_I})E_+\right),\\
c^\prime&=-\frac{1}{8\pi G}\left(\frac{1}{\sqrt{2}}(E_++E_-)\tan(\sqrt{2\sigma_I\bar{\sigma}_I})-\frac{\I}{2\gamma}\sh(4\sqrt{\sigma_I\bar{\sigma}_I})E_+\right)\E^{\I\phi_I}.
\end{align}

The third and forth term in equation \eref{ThetaP1} are redundant. To simplify the expression further, we absorb them back into a redefinition of the canonical variables. Consider $U(1)$ rotation with yet unspecified angle dependent gauge parameter $\psi=\psi(z,\bar{z})$, define
\begin{equation}
U_-=\E^{\psi J}S_-,
\end{equation}
We can then also set
\begin{equation}
\Pi=\E^{\psi J}\Pi'\E^{-\psi J}=J L+ c\bar{X}+\bar{c} X,
\end{equation}
where
\begin{equation}
c=-\frac{1}{8\pi G}\left(\frac{1}{\sqrt{2}}(E_++E_-)\tan(\sqrt{2\sigma_I\bar{\sigma}_I})-\frac{\I}{2\gamma}\sh(4\sqrt{\sigma_I\bar{\sigma}_I})E_+\right)\E^{\I(\phi_I+2\psi)}.\label{cdef}
\end{equation}
The $SL(2,\R)$ symplectic potential transforms according to
\begin{equation}
\operatorname{Tr}\left(\Pi^\prime\bbvar{D}S_-S_-^{-1}\right)=2L\bbvar{D}\psi+\operatorname{Tr}\left(\Pi\bbvar{D}U_-U_-^{-1}\right).
\end{equation}
If we now compare this equation with \eref{ThetaP1}, it is clear that we can cancel the third and forth terms in the expression for the symplectic potential \eref{ThetaP1} by making the following choice\footnote{Notice that in standard Bondi coordintes at asymptotic null infinity $\mathcal{I}^\pm$, the asymptotic time translation can be chosen such that it agrees to leading order in an $1/r$ expansion with our $\mathcal{U}$ coordinate. The shear of such a pulse will then scale within its support  as $\sigma_I(z,\bar{z})=\dot{\sigma}^{(0)}(z,\bar{z})/r+\mathcal{O}(r^{-2})$, hence $\psi=\mathcal{O}(r^{-1})$ as $r\rightarrow \infty$. This is very good, because otherwise our canonical transformation would be inadequate for establishing a connection to perturbative gravity at $\mathcal{I}^\pm$. \label{shearnote}}
\begin{equation}
\psi=\gamma\ln\left[\frac{\tan(\sqrt{2\sigma_I\bar{\sigma}_I})}{\sqrt{2\sigma_I\bar{\sigma}_I}}\right],
\end{equation}
In fact, we obtain
\begin{align}
\Theta^{\mtext{pulse}}_{\mathcal{N}}&=+\frac{1}{16\pi\gamma G}\int_{\mathcal{C}}d^2v_o E_+\left(\ch\left(4\sqrt{\sigma_I\bar{\sigma}_I}\right)+1\right)\bbvar{D}\left[\Delta_++2\gamma\ln\left(\cos(\sqrt{2\sigma_I\bar{\sigma}_I})\right)\right]+\nonumber\\
&\quad-\frac{1}{16\pi\gamma G}\int_{\mathcal{C}}d^2v_oE_+\left(\ch\left(4\sqrt{\sigma_I\bar{\sigma}_I}\right)-1\right)\bbvar{D}\left[\Delta_++\phi_I+2\gamma\ln\left(\frac{\sin(\sqrt{2\sigma_I\bar{\sigma}_I})}{\sqrt{2\sigma_I\bar{\sigma}_I}}\right)\right]+\nonumber\\
&\quad+\int_{\mathcal{C}}d^2v_o\mathrm{Tr}\left(\Pi\bbvar{D}U_-U_-^{-1}\right).\label{ThetaP2}
\end{align}

We are not completely finished yet. The variables in the first two lines are not independent from the variables in the third line. To identify the physical phase, we enlarge the phase space and impose a constraint, which is second class, but not very difficult to solve. Consider first the following angle dependent oscillator type variables
\begin{align}
a&=\frac{1}{\sqrt{8\pi\gamma G}}\sqrt{E_+}\ch\left(2\sqrt{\sigma_I\bar{\sigma}_I}\right)\E^{-\I\left(\Delta_++2\gamma\ln\left(\cos(\sqrt{2\sigma_I\bar{\sigma}_I})\right)\right)},\label{adef}\\
b&=\frac{1}{\sqrt{8\pi\gamma G}}\sqrt{E_+}\sh\left(2\sqrt{\sigma_I\bar{\sigma}_I}\right)\E^{+\I\left(\Delta_++\phi_I+2\gamma\ln\left(\frac{\sin\left(\sqrt{2\sigma_I\bar{\sigma}_I}\right)}{\sqrt{2\sigma_I\bar{\sigma}_I}}\right)\right)}.\label{bdef}
\end{align}
On the enlarged phase space, these oscillators satisfy standard commutation relations for creation and annihilation operators on the cross section,
\begin{equation}
\{a(z,\bar{z}),\bar{a}(z',\bar{z}')\}=\{b(z,\bar{z}),\bar{b}(z',\bar{z}')\}=\I\,\delta^{(2)}_{\mathcal{C}}(z,\bar{z}|z',\bar{z}').
\end{equation}
In addition, we have the constraint
\begin{equation}
a\bar{a}-b\bar{b}=\frac{E_+}{8\pi\gamma G}> 0.\label{abrange}
\end{equation}
We then immediately also find
\begin{align}
E_++E_-&=16\pi\gamma G(L+a\bar{a}),\\
E_--E_+&=16\pi\gamma G(L+b\bar{b}),\\
\sqrt{\sigma_I\bar{\sigma}_I}&=\frac{1}{4}\ln\left(\frac{\sqrt{\bar{a}a}+\sqrt{\bar{b}b}}{\sqrt{\bar{a}a}-\sqrt{\bar{b}b}}\right).
\end{align}
We thus see that the initial and final area densities $E_\pm$ and the shear of the pulse can be all expressed in terms of harmonic oscillator variables. Yet, these oscillators by themselves are \emph{not} Dirac osbervables. Going back to equation \eref{cdef}, in which we defined the $SL(2,\R)$ translation generator $c$, which is a ladder operator in the quantum theory, we  obtain a constraint on the now enlarged phase space. First of all, we note
\begin{equation}
ab=\frac{1}{16\pi\gamma G}E_+\sh\left(4\sqrt{\sigma_I\bar{\sigma}_I}\right)\E^{\I\phi_I+2\I\psi}.
\end{equation}
A short calculation gives the constraint
\begin{equation}
\boxed{c\,\bar{a}\,\bar{b}=
-\gamma\left(L+\bar{a}a\right)\sqrt{2\,\bar{a}a\,\bar{b}b}\tan\left(\frac{1}{2\sqrt{2}}\ln\left(\frac{\sqrt{\bar{a}a}+\sqrt{\bar{b}b}}{\sqrt{\bar{a}a}-\sqrt{\bar{b}b}}\right)\right)+\I\,\bar{a}a\,\bar{b}b.}\label{recrel}
\end{equation}
The Poisson commutation relations for $T^\ast SL(2,\R)$ tell us that this constraint, which commutes with the $SL(2,\R)$ Casimir, forms a second class system with its complex conjugate: the right hand side clearly commutes with its complex conjugate, the left hand side does not.  At the quantum level, we can only impose one of them, the other will map any physical state into its orthogonal complement. In fact, the constraint \eref{recrel} imposes a simple recursion relation among the coefficients in a standard oscillator basis expansion of physical states. Furthermore, if we restrict ourselves to the discrete series representations of $SL(2,\R)$, the constraint can be solved with remarkable ease.  This will be the subject of the next and final subsection.
\subsection{Solving the residual second-class constraint at the quantum level}\label{sec4.2}
\noindent In this section, we show how to solve the constraint \eref{recrel} in the specific case of the unitary discrete series representations of $SL(2,\R)$, in which $L>0$. Following earlier results on loop quantum gravity and isolated horizons \cite{rovelli, thiemann,Thiemann:1997rq,Ashtekar:2000eq,Ashtekar:2004aa,BarberoG.:2012ae,LOSTtheorem}, we use a polymer quantization, in which all geometry is excited at a finite number of topological defects at the null surface. This is achieved by introducing a tessellation of the cross section into a number of plaquettes $\{\square_i\}$, $\bigcup_i\square_i=\mathcal{C}_i$. In each plaquette, we define oscillators and $SL(2,\R)$ generators
\begin{align}
a_i=\int_{\square_i}d^2v_o\,a,\quad b_i=\int_{\square_i}d^2v_o\,b,\quad c_i=\int_{\square_i}d^2v_o\,c,\quad L_i= \int_{\square_i}d^2v_o\,L.
\end{align}
The only non-vanishing commutators are
\begin{align}
[a_i,a^\dagger_j]&=[b_i,b^\dagger_j]=\delta_{ij},\\
[L_i,c_j]&=-\delta_{ij}c_j,\quad[L_i,c_j^\dagger]=+\delta_{ij}c_j^\dagger,\\
[c_i,c_j^\dagger]&=2\delta_{ij}L_i.
\end{align}
Consider then the following basis states for a single plaquette
\begin{equation}
|N,n_a,n_b,n_c\rangle,\label{statevec}
\end{equation}
where $n_a$ and $n_b$ are the occupation numbers of the two oscillators, whereas  $n_c$ is the magnetic quantum number, i.e.\ the eigenvalue of $L$.\footnote{To be precise, there is also a second such magnetic quantum number, which represents the second pair of oscillators $(\utilde{\pi}^A,\utilde{\omega}^A)$ that we introduced in \eref{leftspin}. As far as the solution of the constraint \eref{recrel} is concerned, the corresponding magnetic quantum number $\utilde{n}_c$ is a mere spectator, because the oscillators $(\utilde{\pi}^A,\utilde{\omega}^A)$ commute with the constraint. Hence it is an additional Dirac observable, that encodes the $SL(2,\R)$ shape degrees of freedom of the intrinsic metric $q_{ab}$ on the null cone.}
To satisfy the constraint \eref{abrange}, we restrict ourselves to quantum numbers $n_a$ and $n_b$ such that
\begin{equation}
n_a>n_b.
\end{equation}
In addition, $N$ determines the eigenvalue of the Casimir operator $\hat{Q}=L^2-\frac{1}{2}(cc^\dagger+c^\dagger c)$. For the discrete representations of $SL(2,\R)$, we have $2|n_c|=N,N+1,\dots$ for $N=1,2,\dots$ and
\begin{equation}
\hat{Q}|N,n_a,n_b,n_c\rangle=\frac{1}{4}N(N-2)|N,n_+,n_b,n_c\rangle,\quad
\end{equation}

 Clearly, $N$ and $n_+=n_a-n_b$ are Dirac observables, because the corresponding operators $\hat{Q}$ and $a^\dagger a-b^\dagger b$ commute with the constraint \eref{recrel}. For the same reason, any shift operator that changes the values of $N$ and $n_+$ will be a Dirac observable as well. On the other hand, neither $L$ nor $b^\dagger b$ and $a^\dagger a$ can be Dirac observables. However, the sum $n_-=2L+a^\dagger a+b^\dagger b$ is and so are the shift operators that change it. Consider, for example, a case in which $n_--n_+=2(N+m)>0$ (i.e.\ $E_--E_+>0$). We can now immediately find a unique superposition of states
\begin{equation}
|N,n_+,0,N+m\rangle, |N,n_++1,1,N+m-1\rangle, \dots, |N,n_++m,m,N\rangle\label{spanstates}
\end{equation}
such that the constraint \eref{recrel} is satisfied. In here, we chose an operator ordering such that
\begin{align}
c\,{a}^\dagger{b}^\dagger=
-\gamma\left(L+{a}^\dagger a\right)\sqrt{2\,{a}^\dagger a\,{b}^\dagger b}\tan\left(\frac{1}{2\sqrt{2}}\ln\left(\frac{\sqrt{{a}^\dagger a}+\sqrt{{b}^\dagger b}}{\sqrt{{a}^\dagger a}-\sqrt{{b}^\dagger b}}\right)\right)+\I\,{a}^\dagger a\,{b}^\dagger b\label{qrecrel}
\end{align}
If we try to build physical states in which $n_--n_+<0$ and $L<0$, the recursion relations never terminate. The corresponding physical states contain infinitely many kinematical state vectors \eref{statevec}. At this stage it is an open question whether such states are normalizable kets or  should be rather seen as distributions on the kinematical Hilbert space, i.e.\ elements of the algebraic dual of the kinematical Hilbert space.
 

In defining the constraint \eref{qrecrel}, there are operator ordering ambiguities, but they are mild. The only condition is that the right hand side of the constraint \eref{qrecrel} annihilates the kinematical state $|N,n_a,n_b=0,n_c\rangle$. What is the physical interpretation of the quantum numbers for the Dirac observables we found? A short calculation gives
\begin{align}
&L+a^\dagger a\sim \frac{E_++E_-}{16\pi\gamma G},\label{aainterpret}\\
&L+b^\dagger b\sim\frac{E_--E_+}{16\pi\gamma G},\label{bbLinterpret}\\
&\hat{Q}\sim
\frac{1}{(8\pi G)^2}\left[\frac{1}{4\gamma^2}(E_+-E_-)^2-\frac{1}{\gamma^2}E_+E_-\sh^2(2\sqrt{\sigma\bar{\sigma}})-\frac{1}{2}(E_++E_-)^2\tan^2(\sqrt{2\sigma\bar{\sigma}})\right],\label{Qinterpret}
\end{align}
where $\sim$ has to be understood as equality in the semiclassical limit. In here, $\gamma$ is the Barbero--Immirzi parameter, $E_+$ ($E_-$) is the {final} (initial) area density at $\mathcal{C}_\pm$, and $|\sigma|=|\sigma_I|$ is the norm of the shear in the chosen gauge, see \eref{Ugauge}.\smallskip

A few final remarks.  Throughout this research, we restricted ourselves to discrete but unitary representations of $SL(2,\R)$. We do not know at this stage whether it is possible to find solutions for the continuous series representations on the Hilbert space as well. More work is needed, but we believe that there are no normalizable states in the kernel of \eref{qrecrel}, for which $\hat{Q}=L^2-\tfrac{1}{2}(c^\dagger c+cc^\dagger)<0$. In any case,  states in which $\hat{Q}<0$ will have very little to do with gravity at low energy. This can be argued as follows. If we take a pulse of radiation of finite duration at future (past) null infinity and insert our parametrization back into \eref{Qinterpret}, we find that $\hat{Q}$ is positive until it reaches a critical value, which corresponds to a Bondi news flux of the order of the Planck power $P_{\mtext{Planck}}\approx 1/G$. We will explain this point more carefully elsewhere, but there seems to be a cutoff for the flux. What is also interesting is that for any of our states created from \eref{spanstates}, the operator $L+b^\dagger b$ is always positive definite. This follows immediately from the representation theory of $SL(2,\R)$. Geometrically, $L+b^\dagger b=(E_--E_+)/(16\pi\gamma G)$, which is the difference between the eigenvalues  of area at the final cross section $\mathcal{C}_+$ and the initial cross section $\mathcal{C}_-$. Thus, $E_-\geq E_+$, which is a version of the quasi-local second law at the full non-perturbative quantum level. Throughout this entire work, we assumed $E_\pm>0$. This is perhaps a limitation. Negative values of the area density describe the opposite orientation, i.e.\ an expanding rather than contracting null surface. 
In this case, all creation and annihilation operators are simply exchanged, sending $(a,b,c,L)$ into $(a^\dagger,b^\dagger,c^\dagger,-L)$.

\section{Summary and Interpretation of Results}
\noindent To summarize, we established a non-perturbative quantization of impulsive radiative data on the null cone. At the two cross sections, where the pulses start or terminate, the area density is a Dirac observable. Its spectrum is discrete.
The entire quantization is based on the representation theory of $SL(2,\R)$. For the discrete unitary representations, the strength of the shear degrees of freedom has a discrete spectrum as well. The spectrum is complicated, but it can be inferred, in principle, from \eref{Qinterpret}. At the end of the previous section, we made a  basic conjecture. What makes the discrete series representations stand out in our model is that the recursion relations \eref{qrecrel} terminate after finitely many steps. For the continuous series representations no such mechanism is available. Instead, the coefficients oscillate violently and it seems hard to find finite norm states that lie in the kernel of \eref{qrecrel}. Similar arguments suggest a universal cutoff for the total radiated power as measured at a cross section of future null infinity, but more work is required to make this second statement precise. If true, it would suggest a fundamental UV cutoff due to the fundamental quantum discreteness of geometry.\smallskip

The results where established in three steps: \emph{First}, we studied the {null boundary problem} for general relativity with a parity odd Holst term added to the action in the bulk. The strength of this term relative to the usual Einstein--Hilbert Lagrangian is determined by the Barbero--Immirzi parameter $\gamma$. \emph{Second}, we made a truncation. We performed a gauge fixing and solved the residual constraints, namely the Raychaudhuri constraint and the transport equation for the $SL(2,\R)$ holonomy, for specific impulsive data. By inserting the solution back into the pre-symplectic potential, we equipped the submanifold of impulsive radiative data with a natural symplectic structure inherited from the vast phase space of general relativity. \emph{Third}, we established a quantum representation of the canonical commutation relations. For each null generator, we found three quantum numbers that label physical states: the area of the initial and final cross section and the $SL(2,\R)$ Casimir, which is a combination of the area of the cross sections and the shear of the null generators, see \eref{Qinterpret}. \smallskip

When $\gamma\rightarrow\infty$, the action returns the usual Hilbert--Palatini action for general relativity in the bulk. The addition of the Holst term does not change the classical field equations. However, it affects the algebra of boundary charges. This has observable consequences. The Barbero--Immirzi parameter enters the spectra for the area density and shear degrees of freedom. At the quantum level, the limit $\gamma\rightarrow\infty$ becomes singular. \smallskip

In the previous section, we  studied a very peculiar family of null geometries in which $(E_--E_+)/(16\pi\gamma G)>0$ and $L>0$, representing an infalling quantum null cone.  Clearly, more work is needed to better understand the entire solution space of the recursion relations \eref{qrecrel}. There are numerous more issues that we leave open for future research. For example, it would be very nice to connect our results to the quantum geometry of isolated horizons \cite{Ashtekar:2000eq}. This would be a good test to link the description at the null boundary to a spin network and twisted geometry representation in the bulk \cite{Rovelli:1995ac, thiemann, twist}.1 At this stage, perhaps the most important open technical problem is to take the tensor product of many such {quantum impulsive null initial data} and trace over intermediate edge states to form quantum states for realistic initial data. A related issue is to check whether the gluing procedure is \emph{cylindrical consistent}, i.e.\ stable under refinement \cite{Dittrich:2014ala,LOSTtheorem}. A particular important problem is to define a family of coherent states to take the limit to an asymptotic boundary. In this way, it will be possible to link asymptotic scattering amplitudes at low energies to the full theory at finite distance.

\providecommand{\href}[2]{#2}\begingroup\raggedright\endgroup

\end{document}